\documentclass[5p]{elsarticle}
\usepackage{amsmath}
\usepackage{graphicx}
\usepackage{epstopdf}
\usepackage{amssymb}
\usepackage{wrapfig}
\usepackage[switch,modulo]{lineno}
\usepackage{multicol}
\usepackage{booktabs}
\usepackage{url}
\usepackage{bm}
\usepackage[usenames,dvipsnames]{color}
\usepackage[percent]{overpic}
\usepackage{hyperref}
\usepackage{natbib}
\usepackage{notoccite}
\numberwithin{equation}{section}
\pdfpagewidth=\paperwidth
\pdfpageheight=\paperheight
\usepackage[font=scriptsize]{caption}
\usepackage{lineno}
\usepackage{changes}

\makeatletter

\long\def\@makecaption#1#2{
  \vskip\abovecaptionskip
  \sbox\@tempboxa{{\captionfonts #1: #2}}%
  \ifdim \wd\@tempboxa >\hsize
    {\captionfonts #1: #2\par}
  \else
    \hbox to\hsize{\hfil\box\@tempboxa\hfil}%
  \fi
  \vskip\belowcaptionskip}
\makeatother
\newcommand{\captionfonts}{\scriptsize}

\begin{document}
\begin{frontmatter}

\title{Potential for measuring the longitudinal and lateral profile of muons in TeV air showers with IACTs}

\author[37,42]{A.\,M.\,W.~Mitchell}
\ead{Alison.Mitchell@physik.uzh.ch}
\author[37]{H.\,P.~Dembinski}
\author[37]{R.\,D.~Parsons} 

\address[37]{Max-Planck-Institut f\"{u}r Kernphysik, P.O. Box 103980, D-69029 Heidelberg, Germany}
\address[42]{Physik Institut, Universit\"{a}t Z\"{u}rich, Winterthurerstrasse 190, CH-8057 Z\"{u}rich, Switzerland}

\begin{abstract}

\noindent Muons are copiously produced within hadronic extensive air showers (EAS) occurring in the Earth's atmosphere, and are used by particle air shower detectors as a means of identifying the primary cosmic ray which initiated the EAS. Imaging Atmospheric Cherenkov Telescopes (IACTs), designed for the detection of $\gamma$-ray initiated EAS for the purposes of Very High Energy (VHE) $\gamma$-ray astronomy, are subject to a considerable background signal due to hadronic EAS. Although hadronic EAS are typically rejected for $\gamma$-ray analysis purposes, single muons produced within such showers generate clearly identifiable signals in IACTs and muon images are routinely retained and used for calibration purposes. For IACT arrays operating with a stereoscopic trigger, when a muon triggers one telescope, other telescopes in IACT arrays usually detect the associated hadronic EAS.
We demonstrate for the first time the potential of IACT arrays for competitive measurements of the muon content of air showers, their lateral distribution and longitudinal profile of production slant heights in the TeV energy range. Such information can provide useful input to hadronic interaction models. 

\end{abstract}


\end{frontmatter}

\section{Introduction}
\label{sec:intro}

\noindent Current measurements of the cosmic ray (CR) spectrum and particularly its composition at ultra-high energies (UHE) are dependent on the simulations used to model the development of extensive air showers (EAS) in the atmosphere. The accuracy of ground-based measurements of CRs relies on modelling of the hadronic physics in air showers. Changes in the underlying assumptions of the hadronic interaction models can lead to different interpretations of the measurements, forming a large source of uncertainty \cite{KampertUnger12}.
Aspects where measurements show deviations between the expected air shower development and reality include the muon fraction contained in the air shower as well as the muon lateral distribution (LDF) shape and normalisation \cite{Auger15muonnum,TAmuon}. 

A major open question in CR air shower physics and particle physics is the resolution of this `muon puzzle'; the discrepancies seen between simulations and data in the number of muons produced in hadronic EAS \cite{Auger15muonnum,TAmuon,Pierog17,Petrukhin14}. Indeed, combined results across several experiments confirmed a muon density discrepancy with $8\,\sigma$ significance \citep{Dembinski19}. The current generation of hadronic interaction models used in simulations have been tuned to incorporate measurements from LHC data \citep{Ostapchenko11,Pierog15,Ahn09}. However, the latest models have not been as robustly checked at TeV energies, where a good description of the air shower development is expected. Measurements of the muon content in hadronic EAS at energies below $\sim\,10^{14}$eV in particular, for comparison to the latest LHC-tuned models, are lacking. 

Discrepancies between hadronic interaction models have been largely reduced in the light of LHC data \cite{Pierog17}.
Extensive checking with data from UHE CR experiments such as the Pierre Auger Observatory and Telescope Array has shown that the two experiments are mostly compatible within systematic uncertainties, with both showing a discrepancy to the latest hadronic interaction models \citep{deSouza17,Mallamaci17,Takeishi17}.

Despite recent efforts in improving collider and air shower simulation models, there consistently remains an overabundance of muons in experimental data in comparison to simulations at $\sim\,10^{19}\mathrm{eV}$ \citep{Auger15muonnum,TAmuon}. By contrast, recent results from IceTop at around $10^{14} - 10^{16}\mathrm{eV}$ show no such excess and even a deficit in the number of muons with respect to the latest post-LHC tuned models \citep{icetop18muon}.

Lower energy air showers, in the $10^{12} - 10^{14}\mathrm{eV}$ operating range of IACTs, offer an opportunity to verify whether discrepancies between collider and CR data are connected to the physics of EAS cascade development or arise only in the extrapolation of models to higher energies. Within the sub-PeV energy range in which IACTs operate, the elemental cosmic ray composition is known from satellite and balloon experiments such as AMS-02 and CREAM, providing an opportunity to constrain the cosmic ray composition used as input for simulations \cite{AMSprotonflux15,Kounine2017,Maestro09}.

\begin{figure*}
\centering
\includegraphics[width=1.5\columnwidth]{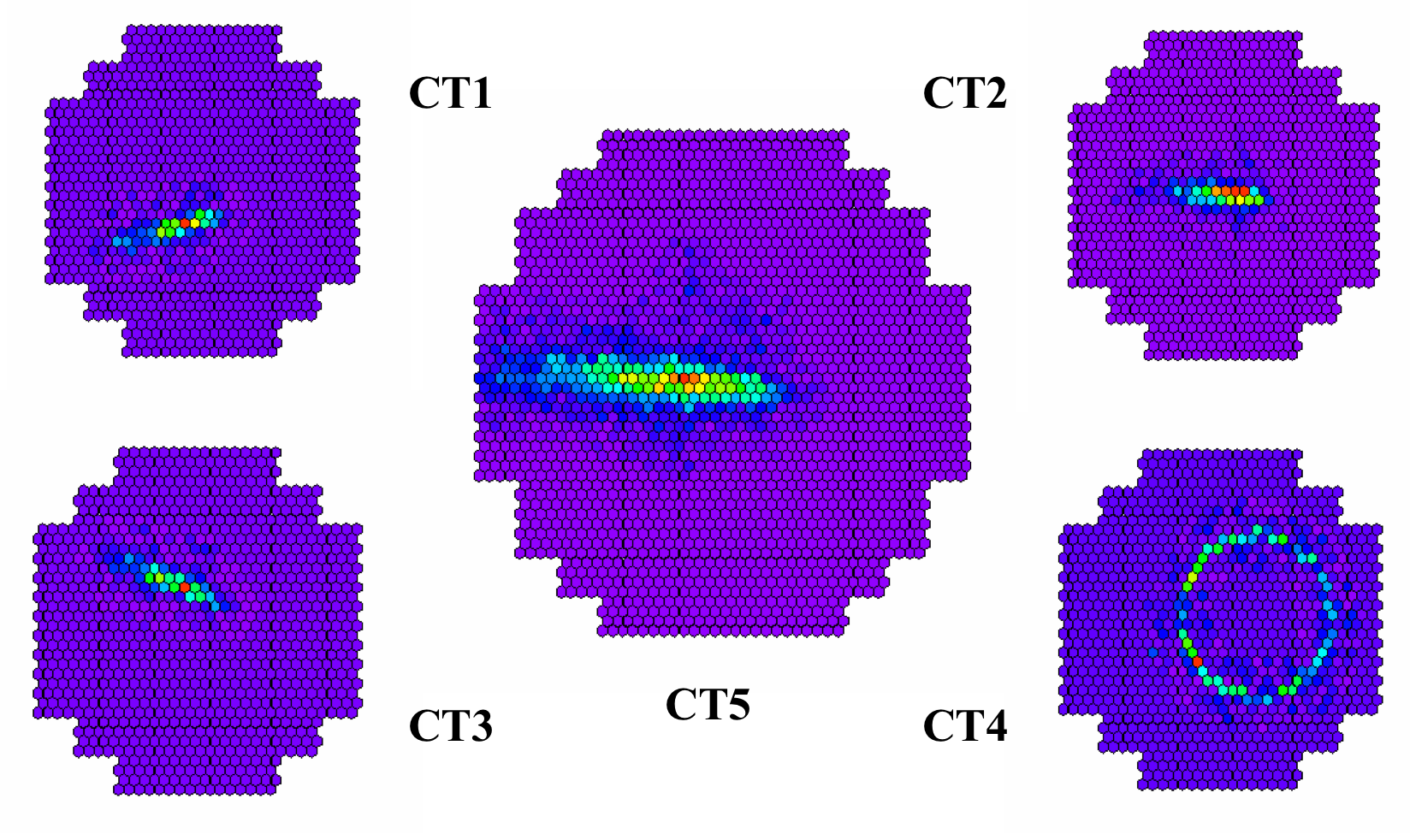}
\caption{Example simulated 10\,TeV proton shower event seen in all 5 H.E.S.S. telescopes. A distinctive muon ring can be seen in CT4, whilst the images in the other telescopes are used to reconstruct the shower core location.}
\label{fig:muonevent}
\end{figure*}

The LHC operates at p-p collision centre-of-mass energies of up to 14\,TeV, corresponding to an initial CR proton energy of $1.0\times10^{17}$\,eV.
Therefore, within the sub-PeV sensitive range of IACT arrays, good agreement with LHC-tuned simulations may be expected, as no extrapolations of hadronic interaction models in energy are required \cite{KampertUnger12}. 

Current IACT arrays have a low effective area to muons; although multiple muons may contribute to the Cherenkov images, typically only one muon per air shower event is identified in the analysis. 
IACTs can identify muons in EAS via their characteristic ring-like images, albeit with a high selection bias towards muons which pass close to or through the telescope mirror dish, thereby generating a more complete ring \citep{Vacanti94}. 
Although in principle ring-shaped images would be caused by any highly energetic charged particle reaching ground level travelling towards the telescope in a straight line, hadronic particles tend to interact more significantly with other particles in the atmosphere, heavier leptons decay more quickly and electrons have lower energies and generally undergo stronger multiple scattering; resulting in energy losses prior to reaching ground level that prohibit the generation of a ring-shaped image.

Muons are usually used by IACT arrays for calibration purposes; as they are minimally ionising and saturate such that above $\sim$\,10~GeV, the same amount of Cherenkov light is produced regardless of the muon energy. Images generated by muons are comprised of Cherenkov emission originating from the last few hundred metres of a muons path prior to passing through (or near to) the mirror dish. As such, muons form a Cherenkov light source of known brightness, that can be used to absolutely calibrate the optical throughput of the Telescopes, with the light undergoing the same path through the last few hundred metres of the atmosphere and through the detector as that of Cherenkov light from $\gamma$-ray showers \cite{Vacanti94}.

The current generation of IACT facilities typically operate stereoscopically, with multiple Telescopes able to simultaneously image the same shower \citep{Hinton04,Lorenz04,Weekes02}. 
In cases where one telescope detects a muon, usually only this image is retained for optical throughput calibration purposes. Additionally, the recorded muon images need to be `clean'; that is, low presence of light from the parent shower in the image such that the ring can be clearly identified. The need for clean images for calibration purposes reduces the overall fraction of muons detected by the standard event selection. In general, however, for any given muon event other nearby IACTs will have detected the associated parent EAS, due to the stereoscopic trigger requirement. An example of such an event including images of the muon and associated shower is shown in Figure \ref{fig:muonevent}. The shower can then be reconstructed using information from the other triggered telescopes, recovering the shower direction and core position within the array, utilising the standard shower reconstruction of these telescopes, although the reconstruction is typically less precise for hadron showers than for $\gamma$-ray showers \citep{Hillas85}. 

\begin{figure}[ht]
\begin{overpic}[width=0.45\textwidth]{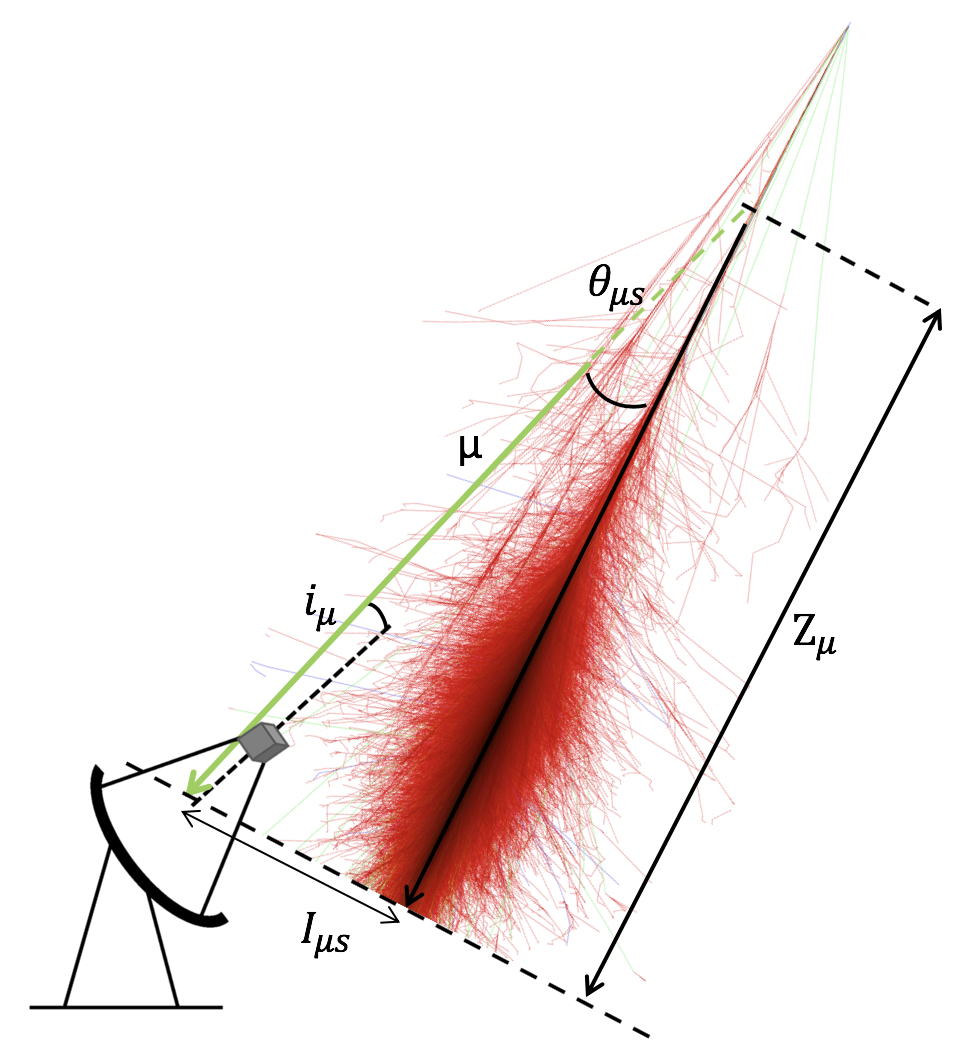}
\end{overpic}
\caption{
Schematic depiction of the geometry of a muon incident on an IACT with the associated 10\,TeV proton shower seen by other telescopes within the array. The inclination angle $i_{\mu}$ of the muon to the telescope axis, orthogonal distance from muon position to shower core position $I_{\mu s}$, muon-shower angle $\theta_{\mu s}$ and muon slant height $Z_{\mu}$ are indicated. Shower image taken from \cite{Corsika17}.}
\label{fig:mugeom}
\end{figure}

Simple geometrical considerations can then be used to find the impact distance $I_{\mu s}$ of the muon to the shower core, as well as the muon production height along the shower axis $Z_{\mu}$, with the same shower geometry as has previously been shown by the Pierre Auger Observatory and Kascade-Grande experiment (see Figure \ref{fig:mugeom}) \citep{Auger14muondepth,Kascade11}.

Measurements of the muon lateral distribution with IACTs extend knowledge of this quantity beyond the results of other experiments such as Kascade and IceCube to the unexplored sub-PeV range \citep{Kascade01,IceCube13}. In comparison to other cosmic ray particle detector experiments, IACTs provide a good directional reconstruction of the cosmic ray shower yet with a poor effective area to muons. 
Nevertheless, the cosmic ray flux is very high at TeV energies, such that the low effective area may be compensated for by the high event statistics \cite{Tanabashi18}.

\section{Simulations}
\label{sec:sims}

\noindent This study was performed using the H.E.S.S. experiment as an example IACT array. To explore the potential for muon measurements with IACTs, we ran a set of dedicated simulations of the H.E.S.S. array with CORSIKA and \textit{sim$\_$telarray}, using the H.E.S.S. analysis software package to produce calibrated images from the simulations \citep{Heck98,Bernlohr08,Aharonian06}.

H.E.S.S. is an array of five IACTs located in the Khomas Highlands of Namibia at 1800~m above sea level \cite{Hinton04}. Originally comprised of four 105~m$^2$ mirror area telescopes operational since 2004, arranged in a square with 100~m sides, the array was enhanced by the addition of a fifth 612~m$^2$ mirror area telescope to the centre of the array in 2012. This makes H.E.S.S. the only currently operational IACT array with telescopes of multiple sizes. 
The increased telescope dish area of the fifth telescope (CT5) lowers the energy threshold of the telescope in comparison to the four smaller telescopes (CT1-4), as the increased area enables more Cherenkov light to be collected per event.

Monoenergetic proton and iron showers were simulated at 10\,TeV and 20$^\circ$ zenith at the H.E.S.S. site, using Corsika version 7.63  and \textit{sim$\_$telarray} version 1.52 compiled with QGSJetII-04 and EPOS~LHC, two hadronic interaction models tuned to LHC data \citep{Ostapchenko06,Ostapchenko11,Pierog15}. Simulations of monoenergetic showers were made in order to generate high statistics for investigating the reconstruction method. Full particle tracking down to ground level enabled the true muon positions at ground to be checked against the reconstructed muons, allowing both the muon identification efficiency and muon purity to be studied. The muon identification efficiency is the fraction of muons passing through the telescope dishes that were correctly identified, whilst the muon purity is the fraction of muon identifications that were truly muons.

Detailed studies of the muon identification efficiency and muon purity were performed using the monoenergetic QGSJetII-04 simulations, with the reduced monoenergetic EPOS~LHC set serving only to compare the true distributions between models. 
Further proton and iron showers were simulated assuming a $E^{-2}$ spectrum using both QGSJetII-04 and EPOS LHC (see Table \ref{tab:sims}), a small sample of which also included full particle tracking  \citep{Ostapchenko11,Pierog15}. 

The number of simulated events corresponds to approximately $\sim\,45$\,minutes of data taking by the H.E.S.S. array. 

\begin{table}
\begin{tabular}{llcl}
Model & Primary & Energy Range & N Showers \\
\hline
QGSJetII-04 & proton & 10\,TeV & $1\times10^6$\\
QGSJetII-04 & iron & 10\,TeV & $8\times10^5$\\
EPOS~LHC & proton & 10\,TeV & $2\times10^4$\\
QGSJetII-04 & proton & 0.8 -- 150\,TeV & $1.2\times10^7$\\
QGSJetII-04 & iron & 2 -- 150\,TeV & $5\times10^6$\\
EPOS~LHC & proton & 0.8 -- 150\,TeV & $1.2\times10^7$\\
EPOS~LHC & iron & 2 -- 150\,TeV & $5\times10^6$\\
\end{tabular}
\caption{CORSIKA simulations used in this study. For the simulations over an energy range, an $E^{-2}$ spectrum was assumed and events were re-weighted in the analysis to reproduce the CR spectrum. The mono-energetic simulations and a small sub-sample of the spectrum simulations included full particle tracking to ground level.}
\label{tab:sims}
\end{table}

\section{Reconstruction method}
\label{sec:reco}

\noindent For the event reconstruction, it is necessary to both identify a muon and reconstruct the associated shower, processes normally executed independently.  
Firstly, muon identification was performed using the procedure outlined in \cite{Mitchell15}, which employs an analytical ring-fitting procedure \cite{Chaudhuri93}.
For each event in which a muon was found, telescopes containing muons were removed from the shower reconstruction. Events containing muons in two telescopes simultaneously were found to occur at a rate $\sim\,1\%$ of the single muon rate in proton initiated showers. 

The shower was reconstructed from images in at least two of the remaining telescopes which detected the shower, using a Hillas parameter based procedure \cite{Hillas85}. 

As an output of the muon reconstruction, the following parameters are obtained: impact position of the muon at ground level with respect to the telescope; inclination angle of the muon relative to the telescope optical axis; ring radius and ring width of the muon ring image produced in the camera \cite{Mitchell15}. The ring radius is determined by the size of the Cherenkov emission cone around the muon, which can be directly related to the muon energy using the standard Cherenkov emission relation:

\begin{equation}
\cos(\theta_c) = 1/(\beta n)~,
\end{equation}

\noindent where $n$ is the refractive index of the atmosphere. Above muon energies of about 10~GeV, this relation tends asymptotically towards muon ring radii of about $1.1^{\circ}$, such that it is difficult to obtain an accurate measure of the muon energy via this relation. For calibration purposes, however, this is an advantageous property of muons seen in IACTs, as the muon energy does not affect the amount of light received by a telescope. Rather, this amount of light is related to the distance across the mirror dish over which light from a muon can be collected and focussed onto the camera, forming a constant quantity \cite{Vacanti94}.

The width of the muon ring is not perfectly sharp, but is broadened by a combination of effects, both physical - changes in the refractive index with wavelength and with height, multiple scattering and ionisation losses - and due to the detector itself - optical point spread function, finite angular pixel size \cite{Vacanti94}.

The muon impact position can be transformed from a position relative to the telescope centre to one relative to the centre of the array, whilst the muon direction can be found from the inclination angle of the muon path to the telescope axis, given by the angular offset of the muon ring centre from the centre of the camera (see Figure \ref{fig:mugeom}) \citep{Vacanti94,Bolz04}.

As part of the shower reconstruction, the shower core position on the ground and the direction of the shower are obtained. The direction of the shower is governed primarily by the orientation of the images in the Cameras, characterised by Hillas parameters \cite{Hillas85}. The energy of the shower can also be reconstructed, governed primarily by the image amplitude, although the shower impact distance and telescope optical throughput must also be taken into account \cite{Hofmann99}. However, since the true shower energy is known for simulations, energy dependent results are shown as a function of true simulated energy, with all results weighted to a E$^{-2.7}$ spectrum.

The distance from the shower core position to the location of the muon gives the muon-shower impact distance $I_{\mu s}$. Single measurements of this quantity over many showers can be used cumulatively to build up the muon lateral distribution. Forming a lateral distribution of muons from a single event is not possible with IACT arrays where $\sim\,1-2$ muons per event are detected. 
Although ground-based UHECR particle detectors such as the Pierre Auger Observatory are capable of reconstructing the LDF on an event-wise basis, measurements are usually made over many events for improved statistics \citep{Auger14muondepth}.

The angle between the reconstructed muon direction and the reconstructed shower direction, $\theta_{\mu s}$, can be used together with the muon-shower impact distance, $I_{\mu s}$, to reconstruct the muon slant height, $Z_\mu$ above ground, as shown schematically in Figure \ref{fig:mugeom} \cite{Auger14muondepth,Kascade11}.

This assumes, however, that the muon and shower axes are in the same projected 2D plane. In 3D space, the muon track and shower axis can be reconstructed, with the point of closest approach as the parameter of interest. An offset from the shower axis is expected given the natural width of hadronic showers (not all particles originate from the primary axis), and that muons may be produced from particle decays with non-negligible transverse momentum ($p_T$) as well as undergo multiple scattering. Neglecting this offset was found by Kascade-Grande to lead to a systematic underestimation of the muon production height by a few percent at most \citep{Kascade11}. We also neglect this offset and consider only the point of closest approach in this study.

\subsection{Cut optimisation}
\label{sec:optimise}

\noindent The images were cleaned according to signal amplitude; all pixels containing at least 10 photoelectrons were kept, along with all neighbouring pixels containing at least 5 photoelectrons (also known as ``tail cut'' cleaning) \cite{Aharonian06}. Pixels with lower amplitudes were cleaned away and not included in the analysis. 

A preliminary selection of shower events for reconstruction was made according to the following criteria: a minimum telescope multiplicity of two; minimum image amplitude in each telescope of 100 photoelectrons; and a minimum of at least 20 pixels in each image after image cleaning. A minimum of at least two telescopes containing regular shower images is required in order to reconstruct a shower stereoscopically. The detection or absence of muon rings in the remaining telescopes does not affect the event selection. With the current analysis, a single telescope image may be used for either shower reconstruction or muon identification, but not both simultaneously.

For the identification of a muon, the selection cuts are based on those used in \cite{Mitchell15}, yet can be relaxed with respect to the requirements for calibration. In this case, the muon reconstruction only needs to be geometrically satisfactory, as the muon image amplitude is not important for our purposes.
Table \ref{tab:muoncuts} summarises the cut parameters used, with the tight values corresponding to the usual values used for calibration and the loose values to the initial cuts applied in this study. The cuts were optimised for muon purity and trigger efficiency. Values for the minimum number of pixels in each image and maximum number of broken and edge pixels (image pixels in the outermost row of camera pixels) remain unchanged with respect to the standard muon calibration procedure and were fixed for each telescope type separately \citep{Mitchell16phd}. The minimum number of pixels in the image and maximum number of broken pixels are generally set to the 3-5\% level, whilst the maximum number of edge pixels is at the $\sim\,1\%$ level.

\begin{figure*}
\includegraphics[width=\columnwidth]{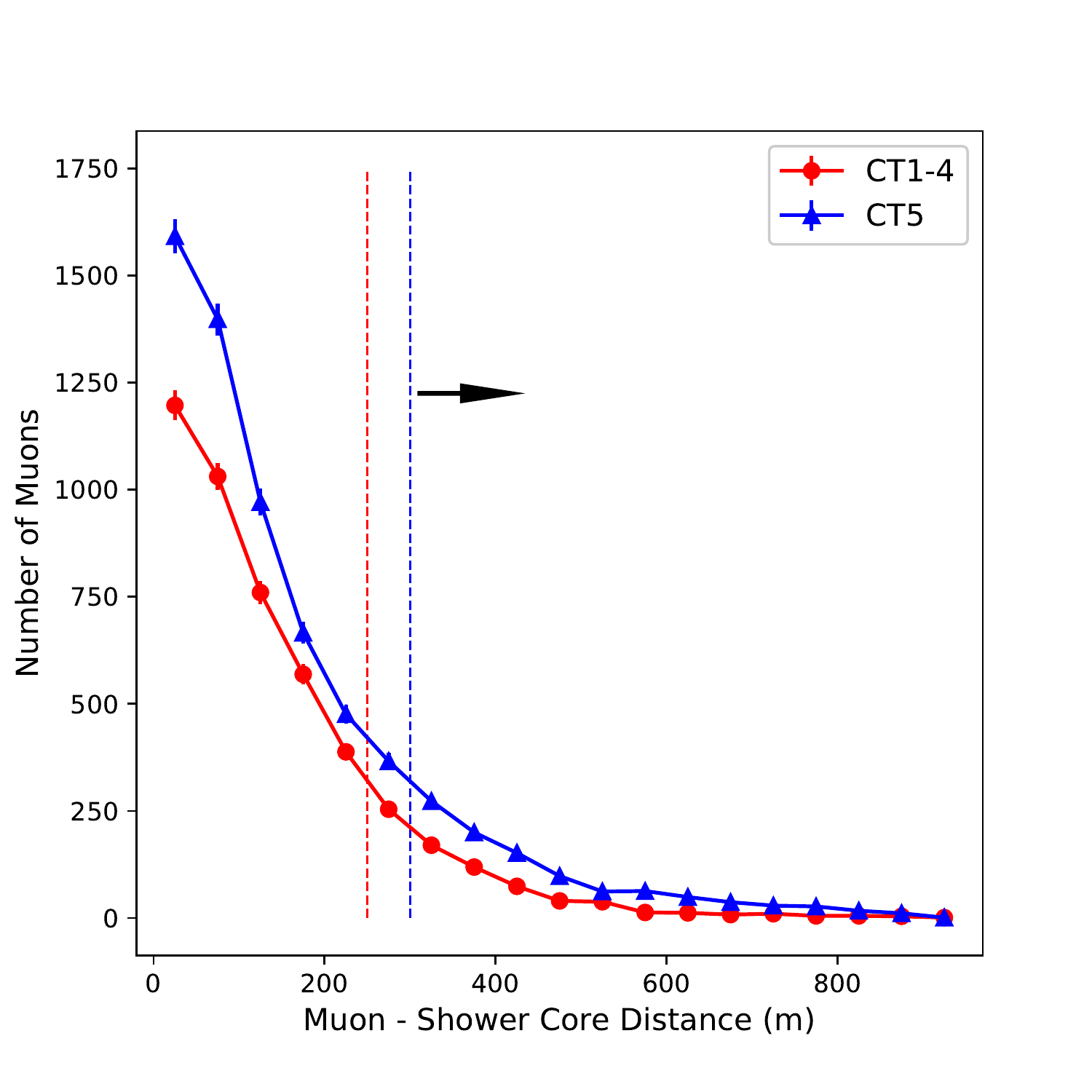}
\includegraphics[width=\columnwidth]{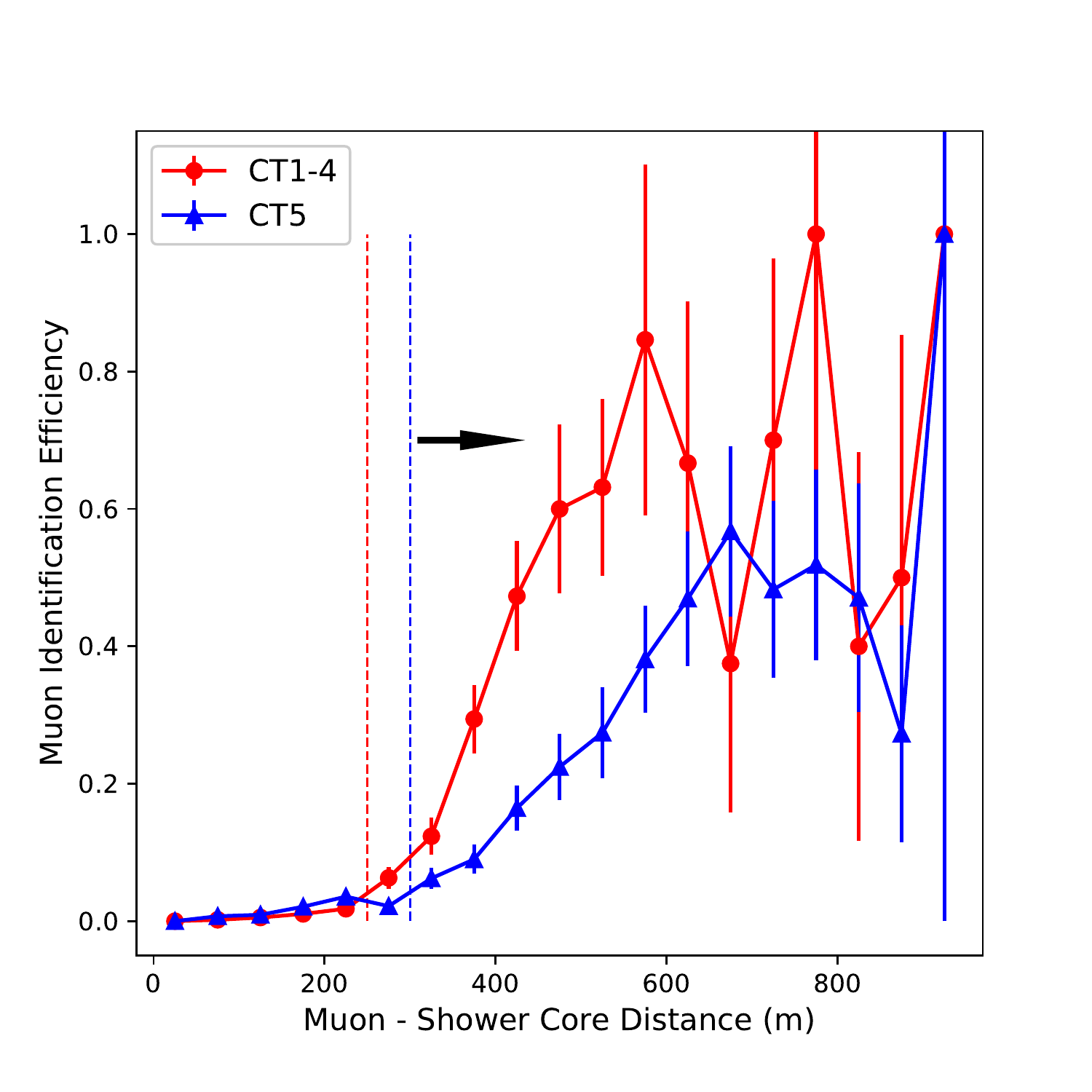}
\caption{Number of muons hitting telescopes (left) and telescope muon identification efficiency (right) as a function of the orthogonal distance of the muon position from the shower core at ground level. The minimum values for CT1-4 and CT5 are given by dashed lines; arrows indicate the regions within which muons are detectable. Error bars are statistical.}
\label{fig:coredisthitcut}
\end{figure*}

\begin{figure*}
\includegraphics[width=\columnwidth]{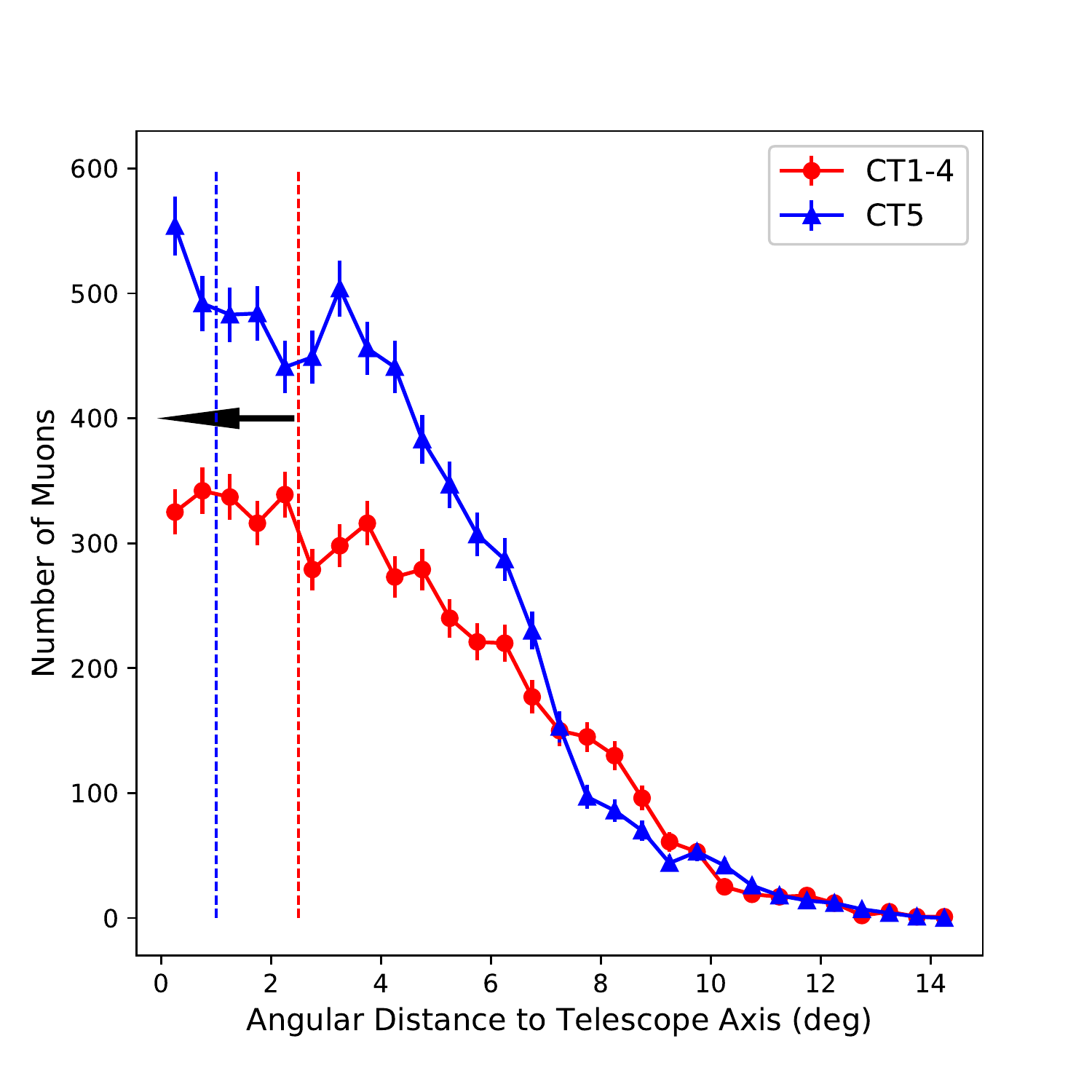}
\includegraphics[width=\columnwidth]{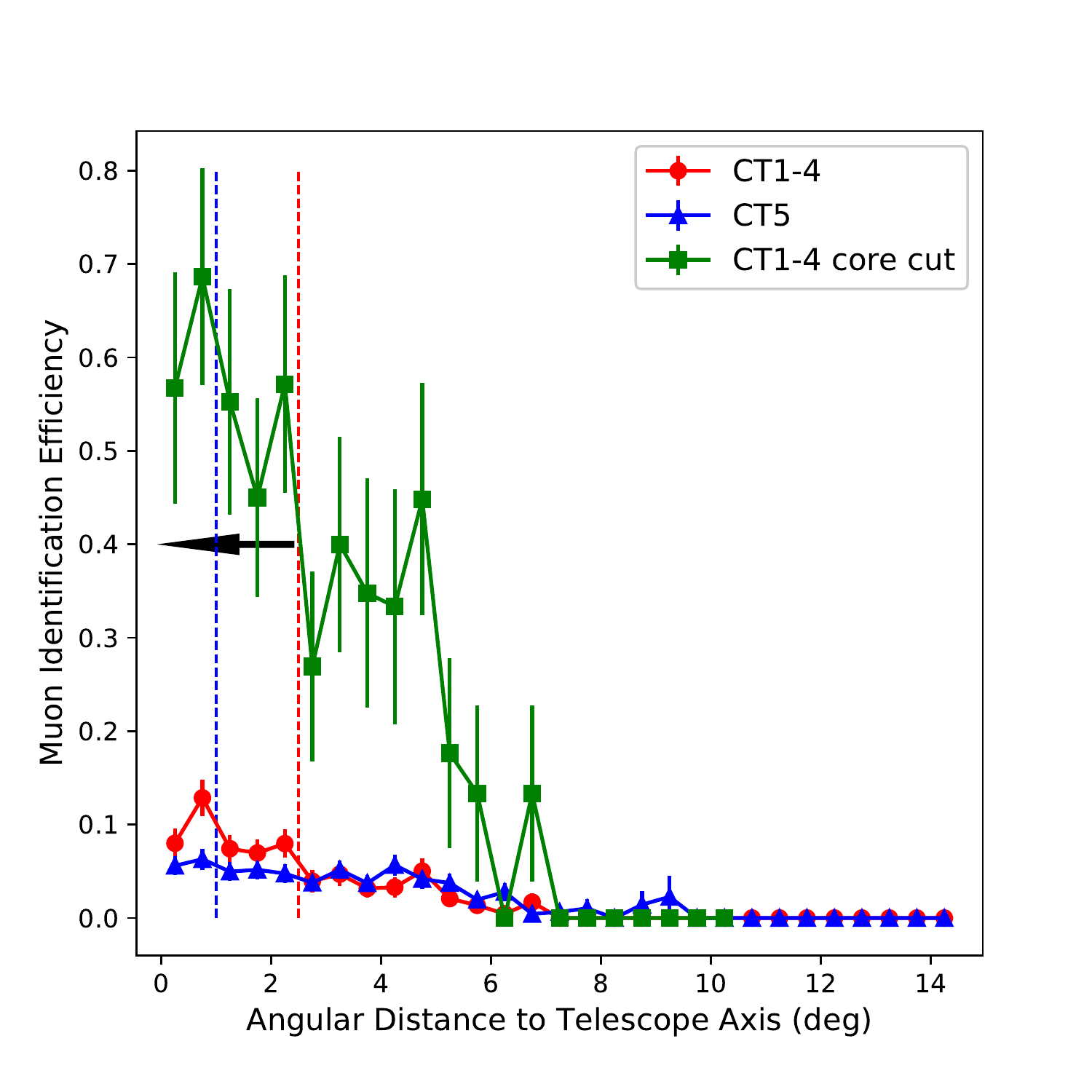}
\caption{Number of muons hitting telescopes (left) and telescope muon identification efficiency (right) as a function of the angular distance of the muon from the telescope optical axis. The maximum values for CT1-4 and CT5 are given by dashed lines; arrows indicate the regions within which muons are detectable. The trigger efficiency is markedly improved by adding a cut on the minimum orthogonal distance between the muon and the shower core location at ground level (green squares in right hand plot), requiring at least 300\,m (see also Figure \ref{fig:coredisthitcut}). Error bars are statistical.}
\label{fig:angdisthitcut}
\end{figure*}

\begin{table*}
\begin{center}
\begin{tabular}{lccc}
 Variable & Loose & Tight & H.E.S.S. standard \\ 
 \hline
 Neighbouring Pixels & $\left< N \right> < 4$ & $\bm{\left<N\right> < 3.5}$ & $\left<N\right> < 3.5$\\
 Ring Completeness & $30\%$ & $\bm{60}$\textbf{\%} & --\\
 Ring Width & $\bm{0.02^\circ-0.2^\circ}$ & $0.04^\circ-0.08^\circ$ & $0.04^\circ-0.08^\circ$ \\
 Impact Parameter & $\bm{0-12}$\,\textbf{m} & $0.9-6.5$\,m & $0.9-6.4$\,m\\
 Ring Radius & $\bm{0.9^\circ-1.5^\circ}$ & $1.0^\circ-1.5^\circ$ & $1.0^\circ-1.5^\circ$\\
 Outer Ring Radius & $\bm{< 3.5^\circ}$ & $< 2.2^\circ$ & $< 2.^\circ$
\end{tabular}
\caption{Selection cut parameters for identifying muon events within H.E.S.S., quoted for CT1-4. The final set of cuts adopted are indicated in bold (see also Figure \ref{fig:muontp}) Cuts may be adapted slightly for CT5, in particular on the impact parameter and outer ring radius (ring radius + muon ring centre offset from camera centre) \cite{Mitchell15}. Note that the ring completeness is not currently used as part of the standard muon analysis within H.E.S.S. for optical efficiency calibration purposes. }
\label{tab:muoncuts}
\end{center}
\end{table*}

\subsection{Muon purity and muon identification efficiency}
\label{sec:puritytrigeff}

\noindent When using the standard H.E.S.S. muon selection cuts for calibration (see Table \ref{tab:muoncuts}), it was found that up to 10\% of the muons detected were in fact false detections, such that the typical muon purity is $\sim\,90\%$. The amount of false detections could be further reduced by the introduction of a ring completeness cut, only accepting muon rings with at least 60\% of the ring present in the image (for CT1-4) and 45\% for CT5. Subsequently, it was found that $\sim\,95\%$ of the muon events surviving these cuts in the test sample with full particle tracking were in fact true detections.
When adding this ring completeness cut to the bulk simulations it was found that the overall number of events passing cuts also decreased by 10\%, consistent with the test sample. Therefore, we can be confident that the muon purity is typically $\gtrsim 90\%$ in the events passing standard muon selection cuts. 

\begin{figure}
\begin{center}
\includegraphics[trim=5cm 5.cm 3.5cm 4.cm,clip,width=0.49\columnwidth]{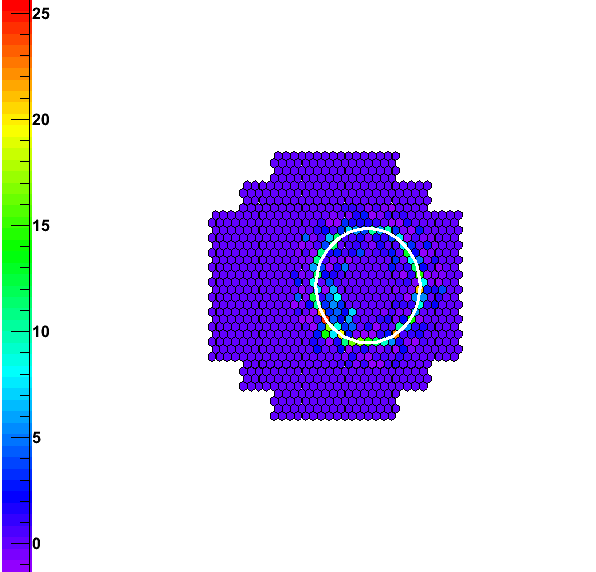}
\begin{overpic}[trim=5cm 5.cm 3.5cm 4.cm,clip,width=0.49\columnwidth]{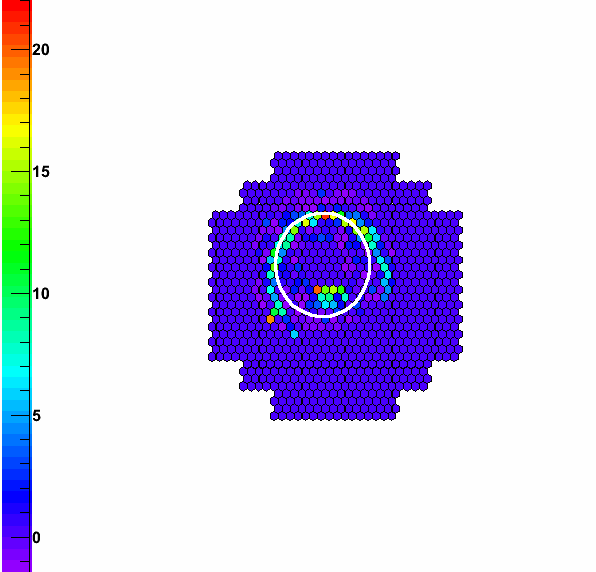}
\end{overpic}
\caption{Example events in which a muon hit the telescope, with the best fit circle is shown in white. Left: a muon identified by the standard muon procedure. Right: a muon event where the fit is biased by the presence of the parent shower in the image, resulting in a circle that does not pass the muon identification cuts.}
\label{fig:unseenmuons}
\end{center}
\end{figure}

In the simulation sample with full particle tracking (in which the position of all particles at ground level was recorded), it was found that muons were only identified if they were on-axis (angular distance within $2^\circ$ of the telescope axis or $1.4^\circ$ for CT5) and if the orthogonal distance from the muon position to the true shower core position at ground level was greater than 250\,m (300\,m for CT5). 
The variation of trigger efficiency and number of muons with muon-shower core distance $I_{\mu s}$ and angular distance is shown in Figures \ref{fig:coredisthitcut} and \ref{fig:angdisthitcut} respectively.

Whilst in Figure \ref{fig:angdisthitcut} both the number of muons and trigger efficiency drop with increasing angular distance, in Figure \ref{fig:coredisthitcut} although the number of muons decreases with increasing shower core distance, the trigger efficiency increases. Note that the fluctuations seen at large core distances in Figure \ref{fig:coredisthitcut} are not significant.

As reflected by the trends shown in Figure \ref{fig:coredisthitcut}, one of the main reasons for non-identification of a muon, is that the image may be confused or dominated by the Cherenkov emission from the rest of the particle shower, such that the muon ring fit is biased by the presence of other emission and cannot be easily identified.
Often, Cherenkov emission from the hadronic shower adds charge in a muon image to the centre of the muon ring. An example of such an event is shown in Figure \ref{fig:unseenmuons}, where the additional emission from the hadronic shower clearly biases the fit to the muon ring. In this case, the ring would also have comparatively low ring completeness (based on presence of charge along the best fit circle), low ring radius and a high average number of neighbouring pixels, thereby failing the selection cuts in Table \ref{tab:muoncuts}.

The trends with angular distance shown in Figure \ref{fig:angdisthitcut} illustrate the fact that the closer a muon direction is to being parallel to the telescope optical axis (more on-axis), the more complete the generated muon ring is, making the muons easier to identify.

The overall telescope acceptance to muons therefore drops off not only with distance of the muon position on the ground from the telescope, but also with angular distance of the muon from the telescope axis. True muon events are expected to trigger a telescope if they pas through the telescope mirror dish; if they have an angular distance of less than $\sim\,2.5^\circ$ to the telescope optical axis and a distance of at least 250\,m to the shower core position, reducing image contamination and potential bias from the parent hadronic shower. 

From the simulations with full particle tracking, it was possible to identify muons that should have been seen by the telescopes, using the aforementioned limits on angular distance and core distance as well as only considering events in which telescopes triggered on the shower. Of all of the possible muons that could have been detected by the telescopes, 45\% were detected. 

\begin{figure}
\includegraphics[width=\columnwidth]{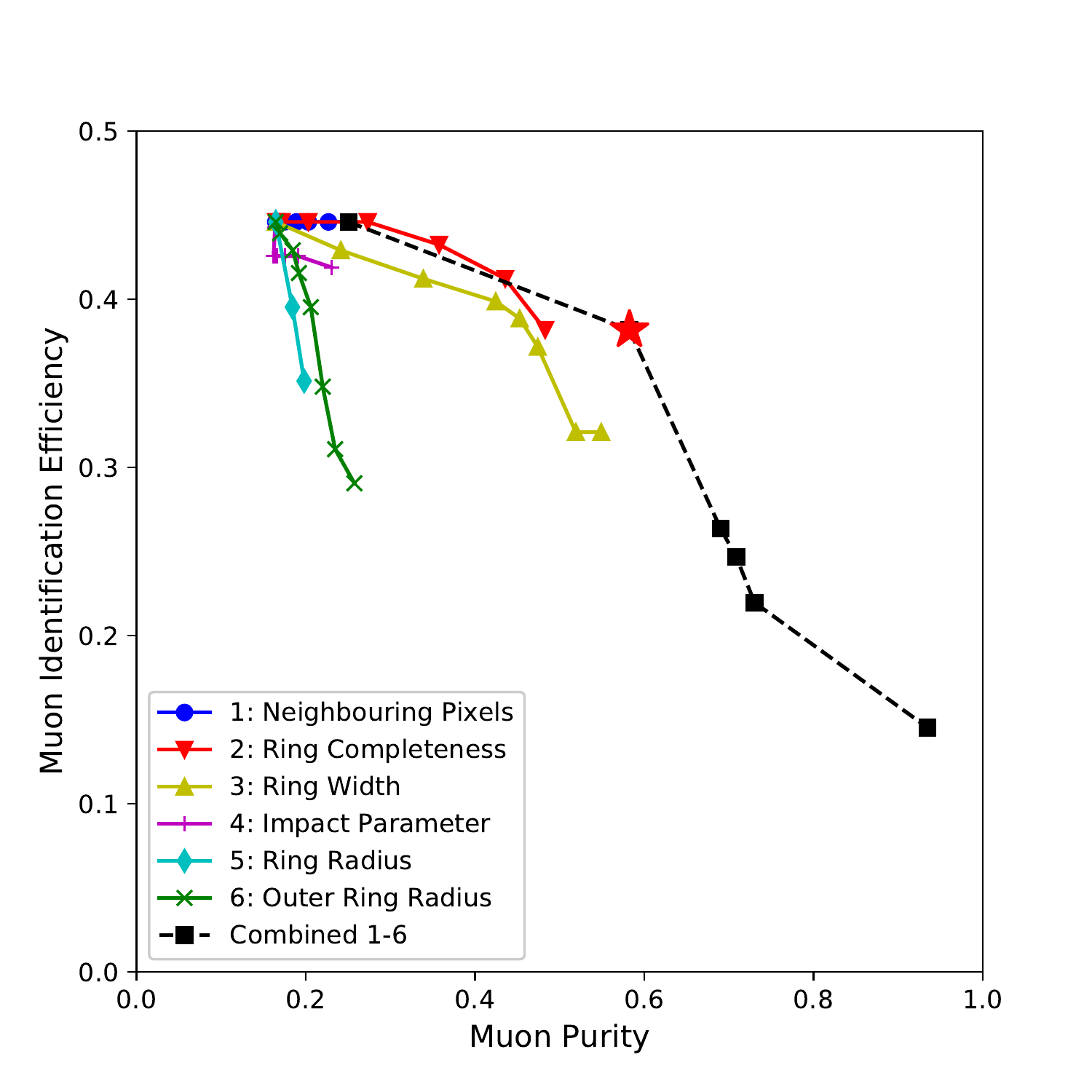}
\caption{The effect of increasing tightness of muon selection cuts from loose to tight (see Table \ref{tab:muoncuts}) on muon identification efficiency and muon purity for CT1-4. The black dashed line indicates the effect of combining cuts 1-6 at their tight values in order, with the final point showing the muon purity usually achieved for calibration purposes ($>90\%$). The chosen cut combination, representing the best trade-off for this study, is indicated by a red star.}
\label{fig:muontp}
\end{figure}

When using the standard cut values (`tight' values in Table \ref{tab:muoncuts}), optimised to identify good muon events for calibration purposes, a muon Identification efficiency of only $\sim\,15\%$ was found, albeit with $93\%$ muon purity when applying the cuts normally used for muon identification (see Figure \ref{fig:muontp}). 
Whilst a pure muon sample is important for optical throughput calibration using muons, this results in a low muon identification efficiency. For the purposes of this analysis, it is desirable to increase the muon identification efficiency, in order to obtain a more representative sample of muon-shower events, for which some muon purity can be compromised.

Figure \ref{fig:muontp} demonstrates the trade-off relationship between muon identification efficiency and muon purity as a function of different cuts on the muon image. In order to test this, all cuts were set to loose ranges; the ratio of the number of true muon events to the total number of events passing cuts provided the muon purity, whilst the muon identification efficiency is given by the fraction of muon events expected to be seen (given the geometry of the shower, see Figures \ref{fig:coredisthitcut} and \ref{fig:angdisthitcut}) that were actually identified. 

For each cut parameter, the range of allowed values was gradually tightened until the standard (tight) cut value was reached, as given in Table \ref{tab:muoncuts}. In all cases, tightening the cuts led to improved muon purity at the cost of a decreased muon identification efficiency, as expected. 
It can be seen that ring completeness is the single most effective cut parameter in improving the muon purity, being the only parameter to achieve a purity of greater than $50\%$ in isolation. Nevertheless, by combining cut parameters (at their tight values in Table \ref{tab:muoncuts}) in order from 1-6 as labelled in Figure \ref{fig:muontp}, it can be seen that with the addition of each new cut parameter, the purity improves. 

By keeping tight cut values for the average number of neighbouring pixels and ring completeness, with loose values for all other cut parameters as listed in Table \ref{tab:muoncuts}, it is possible to increase the muon identification efficiency to 38\%, albeit with a purity reduced from 93\% to 58\%. We adopted this set of cuts (indicated by a red star in Figure \ref{fig:muontp}) as a reasonable trade-off between high muon purity and good muon identification efficiency to mitigate selection bias effects. An alternative set of cuts, including all cuts at tight values in Table \ref{tab:muoncuts} (except for outer ring radius), was tested and the results found to be consistent within errors, yet with reduced statistics.

\subsection{Muon - shower event reconstruction}
\label{sec:mushreco}

\noindent Figure \ref{fig:muonevent} shows an example Monte Carlo event in 5 telescopes, with a muon seen by CT4 and the parent shower seen by the other four telescopes. Despite this proton shower being a hadronic event, as long as the images are approximately elliptical, use of a Hillas parameter based shower reconstruction to identify the shower direction and core location can still be expected to perform reasonably well. The muon ring in CT4 is comparatively clean, that is, the background hadronic shower does not contribute significantly to the image size. In this case, the associated parent shower is not seen in the image. This may be due to the muon being produced at a large core (or angular) distance from the shower axis, such that the shower is not detected by the telescope. Alternatively, the muon may have triggered the Camera at a slightly different time, such as in advance of the rest of the shower, that the charge integrated over the 16\,ns  integration window is not sufficient to also contain the shower \cite{Blake82,Blake90}.

The muon in Figure \ref{fig:muonevent} can be seen to have a ring width of approximately one pixel, whilst muon events in CT5 typically have intrinsic widths covering up to $\sim\,3-5$ pixels. 
This is due to the smaller angular pixel size and from broadening of the muon ring, which occurs due to both optical effects (such as mirror aberration) and natural contributions including multiple scattering, which tends to be more significant for lower energy particles \cite{Vacanti94}.

Whilst the event in Figure \ref{fig:muonevent} is a particularly conducive example for our purposes, there are limitations to treating the hadronic shower as a $\gamma$-ray shower  in the reconstruction. This approach treats all images as if they are elliptical; the greater the deviation from elliptical, the more biased the reconstruction. 

Systematic biases in the energy reconstruction are also to be expected given the differences in how energy is distributed within the hadronic shower during the shower development in comparison to electromagnetic showers, on which the energy reconstruction is based.  

As a Hillas parameter based reconstruction was used \cite{Hillas85}, we applied a simple cut on the level of standard deviation of the Hillas width $w$ and length $l$ between $N$ telescopes; 

\begin{equation}
\sigma_w^2 = \frac{\sum_i^N(w_i - \langle w \rangle )^2}{N}~,
\end{equation}

\noindent and analogously for the length, requiring $\sigma_{w} < 0.06^\circ$. 
This cut restricts the analysis to events with similar impact distances between telescopes and effectively ensures that the main shower (or the same sub-shower) was detected in the images used for the event reconstruction. Removing events with high deviation, it was found that the Pearson correlation coefficient between reconstructed and true shower energy increased from $\sim\,15\%$ without cuts on Hillas width deviation to $\sim\,26\%$ for p and $\sim\,22\%$ for Fe, with an energy bias $(E_{\mathrm{reco}} - E_{\mathrm{true}}) / E_{\mathrm{true}}$ above 10\,TeV of $< 0.3$ for both p and Fe. This is an indication that the selected showers are more $\gamma$-ray like, such that the $\gamma$-ray based shower reconstruction performance improves, albeit with a large reduction in event statistics. 

From the shower sample with fully tracked particles, it was found that the core resolution is $\sim\,60-80$\,m (68\% containment) for showers falling within the array, with the aforementioned cuts on Hillas parameter deviation applied. For comparison, a core reconstruction accuracy of $\sigma_r = 30$\,m was quoted by HEGRA for CR proton shower reconstruction, whilst we obtain a width of $\approx 40 - 50$\,m from a 2D Gaussian fit to proton and iron showers respectively \cite{Hegra99}. Note that this $\sim\,30$\,m accuracy was achieved by \cite{Hegra99} using a cut on the Mean Scaled Width (MSCW) relative to protons; implementing a similar cut may improve the resolution.\footnote{Implementing the necessary changes to the software pipeline to enable such a cut is beyond the scope of this study.} Tightening of the Hillas deviation cut beyond this level or cutting on MSCW relative to $\gamma$-rays was found to make no further improvement on the core or energy reconstruction. 

As expected, the reconstructed shower core position occurs more often within the array than outside. This is partly an artefact of requiring a minimum telescope multiplicity of 3 telescopes triggering on a given event in order for that event to be used in the reconstruction. As such, events falling within the array are more likely to trigger at least 3 telescopes - the true shower distribution was simulated with a $10.^{\circ}$ cone and was more uniformly spread on the ground.

The reconstructed position of the muons is always coincident with the location of one of the five H.E.S.S. telescopes, as the muon is otherwise not identified. This is due to the muon selection cuts requiring that a near complete muon ring is seen by the telescope Camera, such that muons are only recorded with high purity when travelling through the telescope mirror dish. Fewer incident muons were recorded in the central telescope (CT5) than on the smaller telescopes (CT1-4). This can be understood as due to the lower energy threshold of CT5, its smaller field of view and its larger collection area. Whilst CT5 detects more muons overall than CT1-4, for the types of events required in this analysis, it is more likely that CT5 and one other telescope detect the associated shower when a muon is found in one of CT1-4 than the opposite case of an associated shower being detected in at least two of CT1-4 with a clean muon (and minimal shower) detected in CT5. That is due at least in part to the array geometry; with a shower detected in multiple CT1-4 telescopes it is difficult to simultaneously obtain a `clean' muon in CT5. 

\begin{figure}[ht]
\includegraphics[trim=0mm 0mm 5mm 0mm,clip=true,width=\columnwidth]{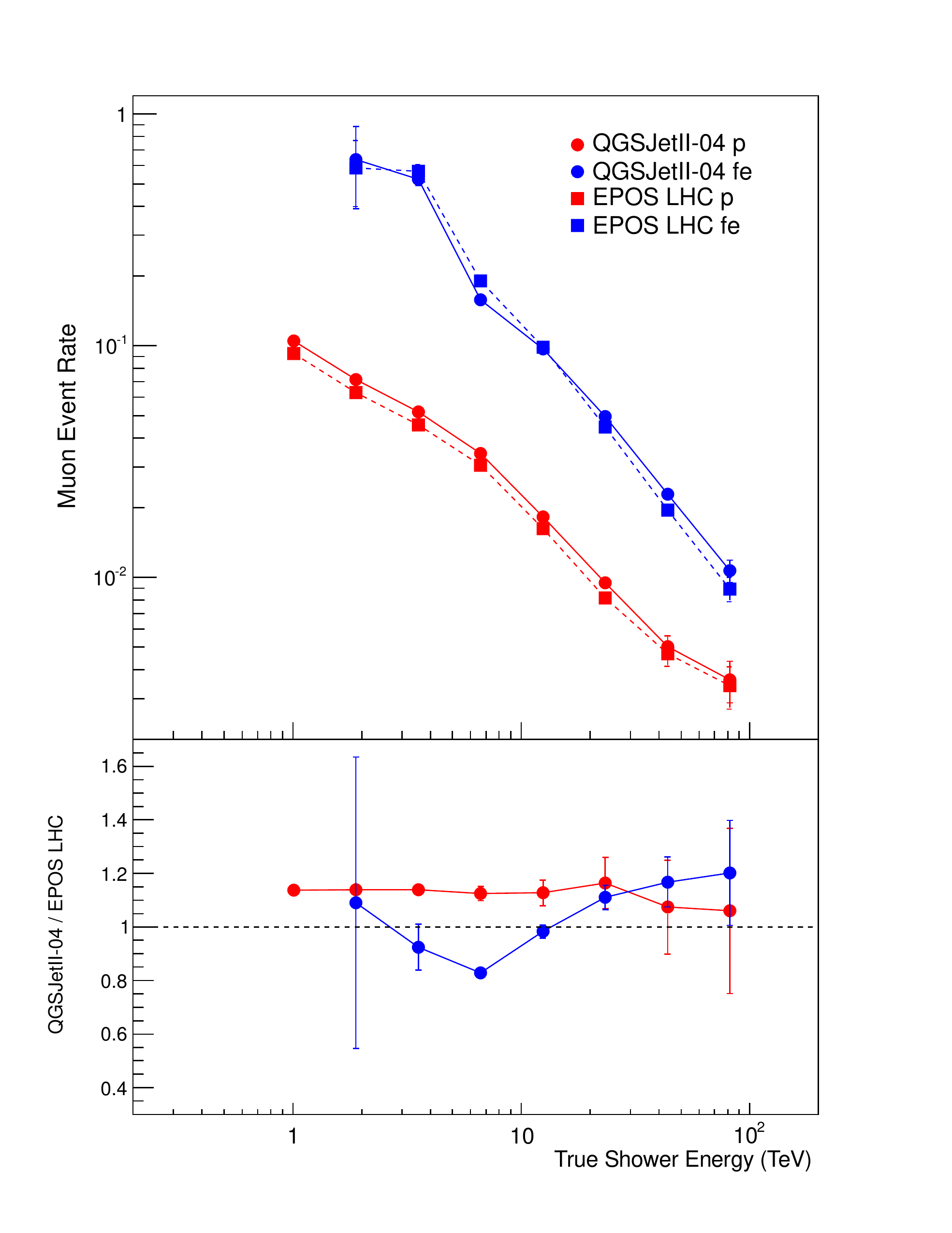}
\caption{Top: Comparison of the muon event rate in showers triggering the array that contain muons between proton and iron primaries for two different hadronic interaction models and as a function of true shower energy, normalised for the same number of showers triggering at least two telescopes. Bottom: Relative muon abundance in QGSJetII-04 with respect to EPOS-LHC for proton and iron showers.}
\label{fig:rateenergy}
\end{figure}

\section{Results}
\label{sec:results}

\subsection{Muon event rate}
\label{sec:rate}

\noindent The muon event rate was determined as the fraction of all showers passing selection cuts (minimum telescope multiplicity and image size) which contained identifiable muons and for which both the muon and associated shower could be reconstructed. For comparison, the rates were normalised for the same total number of showers triggering at least two telescopes per energy bin. 

This muon event rate is shown as a function of true shower energy in Figure \ref{fig:rateenergy}. The muon event rate was found to be higher in iron simulations than in proton simulations as expected, although the ratio of the number of muons in iron showers to that of protons showers is larger than the $\sim\,1.4$ expected, at $\sim\,10$ but decreasing with increasing energy \cite{Dembinski18}. Although the curves are normalised for the same number of showers triggering at least two telescopes, this indicates that the array preferentially triggers on iron showers at a given energy, with the bias decreasing towards higher energies. 

This is likely due to a strong selection effect, whereby the identification of muons is easier for iron initiated showers than proton due to both the intrinsically higher muon content and the increased proportion of sub-showers leading to muons scattering far from the shower core. Figure \ref{fig:coredisthitcut} shows that the muon identification efficiency increases with distance between the muon and the shower core (reduced image contamination); therefore, it will improve with the number of sub-showers, corresponding to the mass of the primary and the shower energy. The decrease in ratio with increasing shower energy is hence also expected.
This selection bias may be improved in future with the inclusion of more robust muon identification algorithms, able to identify partial muon rings in more complex images.

The muon event rate in proton showers was also slightly greater when using QGSJetII-04 than using EPOS-LHC, at the level of $\sim\,15\%$. This is consistent with recent findings that the muon content is rather higher in QGSJetII-04 than expected at TeV energies \citep{Dedenko15}. 

However, for iron showers it is interesting to note that the relative abundance of muons in QGSJetII-04 is consistent with or even lower than EPOS\,LHC until primary energies of about 10\,TeV are reached, at which point the ratio becomes compatible with that seen in proton showers.

\subsection{Muon lateral distribution}
\label{sec:muonldf}

\begin{figure}
\centering
\includegraphics[width=0.48\textwidth]{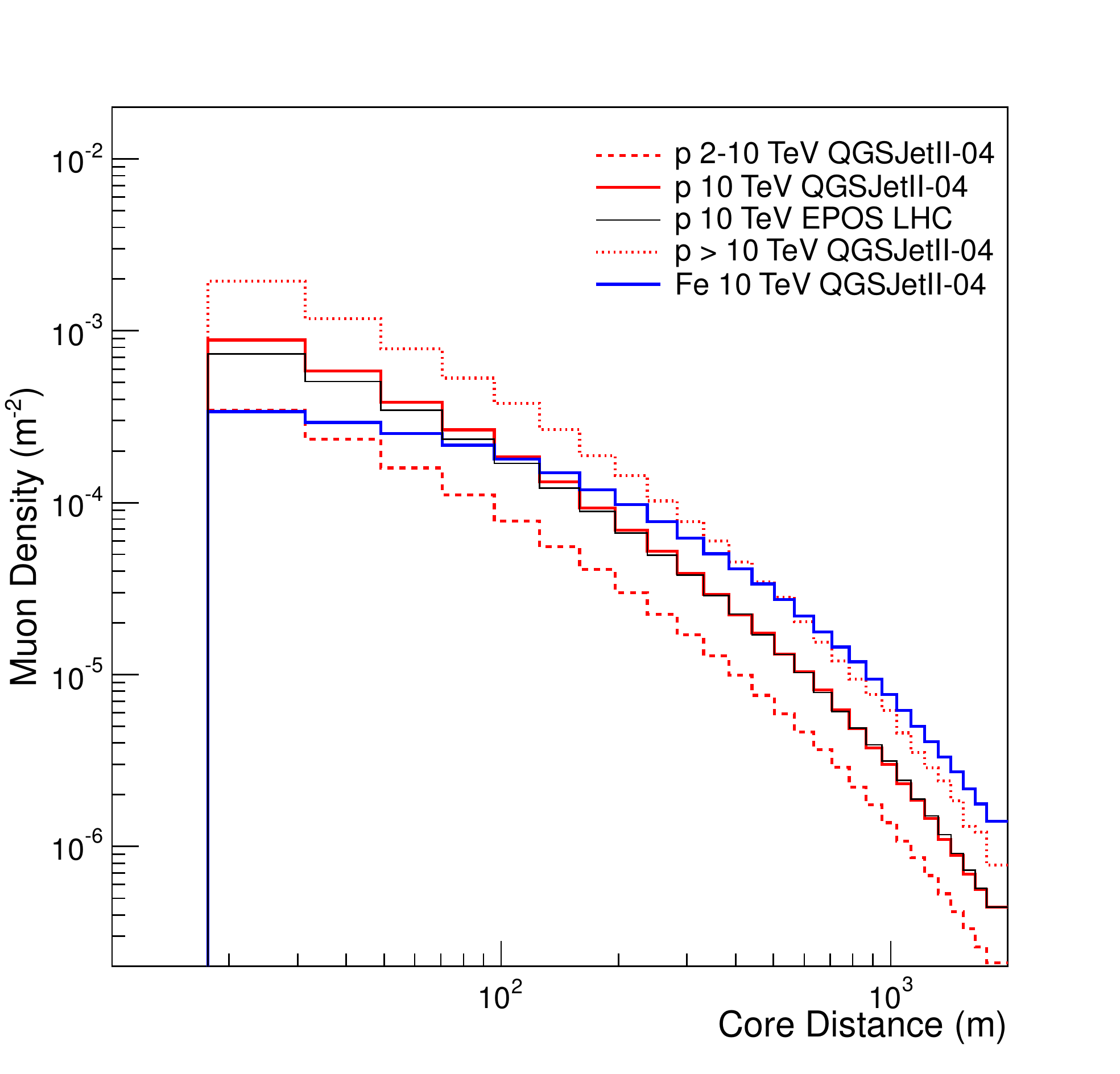}
\caption{True muon lateral distribution at ground using simulations with particle tracking enabled. Solid lines indicate the true distribution for 10\,TeV showers. }
\label{fig:truemuldf}
\end{figure}

The muon lateral distribution is a measure of the muon density $\rho_\mu$ at ground within EAS as a function of distance to the shower core, $r$. Based on a small set of simulations with full particle tracking enabled, the true muon lateral distribution at ground was found by using the expression: 
\begin{equation}
    \rho_\mu = \frac{N_\mu}{2\pi r \mathrm{d}r N_{\mathrm{ev}}}~,
    \label{eq:truemu}
\end{equation}
where $\mathrm{d}r$ is the radial width of the bin at a distance $r$ from the shower core; $N_\mu$ is the number of muons in each radial bin and $N_{\mathrm{ev}}$ is the number of simulated events. The reconstruction was performed in the tilted frame such that a factor of $\cos \theta_z$ (where $\theta_z$ is the zenith angle of the showers, $20^\circ$ in this study) is implicitly taken into account. This quantity is shown in Figure \ref{fig:truemuldf}, where the normalisation of the muon lateral distribution can be seen to increase with increasing shower energy. Curves generated using QGSJetII-04 are shown for monoenergetic simulations of proton and iron at 10\,TeV, as well as for simulations with an $E^{-2}$ spectrum in the energy ranges 2-10\,TeV and 10-150\,TeV. The muon lateral distribution from 10\,TeV iron showers has a flatter shape than that of 10\,TeV proton showers below a few hundred metres. For comparison, the true muon lateral distribution for 10\,TeV proton showers using EPOS\,LHC is also shown, yet only slight differences with respect to QGSJetII-04 at low $r$ are seen.

In order to reconstruct the muon lateral distribution at ground using IACTs, the following expression for the measured muon density $\rho^*_\mu$ at ground was used:
\begin{equation}
    \rho^*_\mu = \frac{N^*_\mu}{A_{\mathrm{eff}}N^*_{\mathrm{ev}}}~,
    \label{eq:recomu}
\end{equation}
where $N^*_\mu$ is the number of detected muons; $N^*_{\mathrm{ev}}$ is the number of events triggering at least two telescopes; and $A_{\mathrm{eff}}$ is the telescope array effective area to muons. 
To avoid a bias in the measurement towards only showers where a muon hit a telescope, the number of events triggering at least two telescopes irrespective of whether or not a muon was detected is used for $N^*_{\mathrm{ev}}$. 
The effective area, which is unknown a priori, was determined for each energy range separately, as well as for proton and iron showers, by setting equations \eqref{eq:truemu} and \eqref{eq:recomu} equal (i.e. $\rho_\mu = \rho^*_\mu$ by definition) and solving for $A_{\mathrm{eff}}$. 

\begin{figure*}
\includegraphics[width=0.48\textwidth]{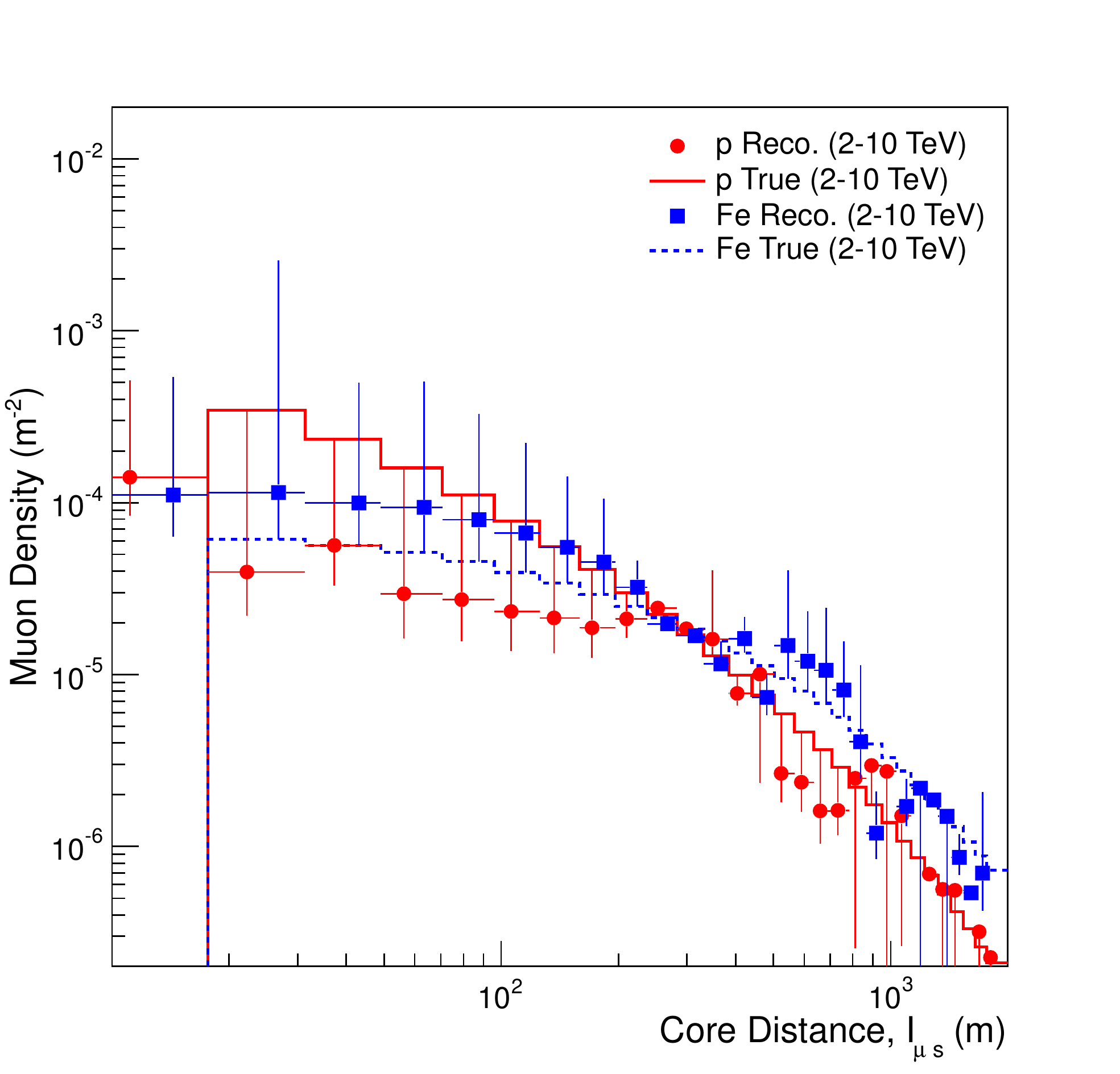}
\includegraphics[width=0.48\textwidth]{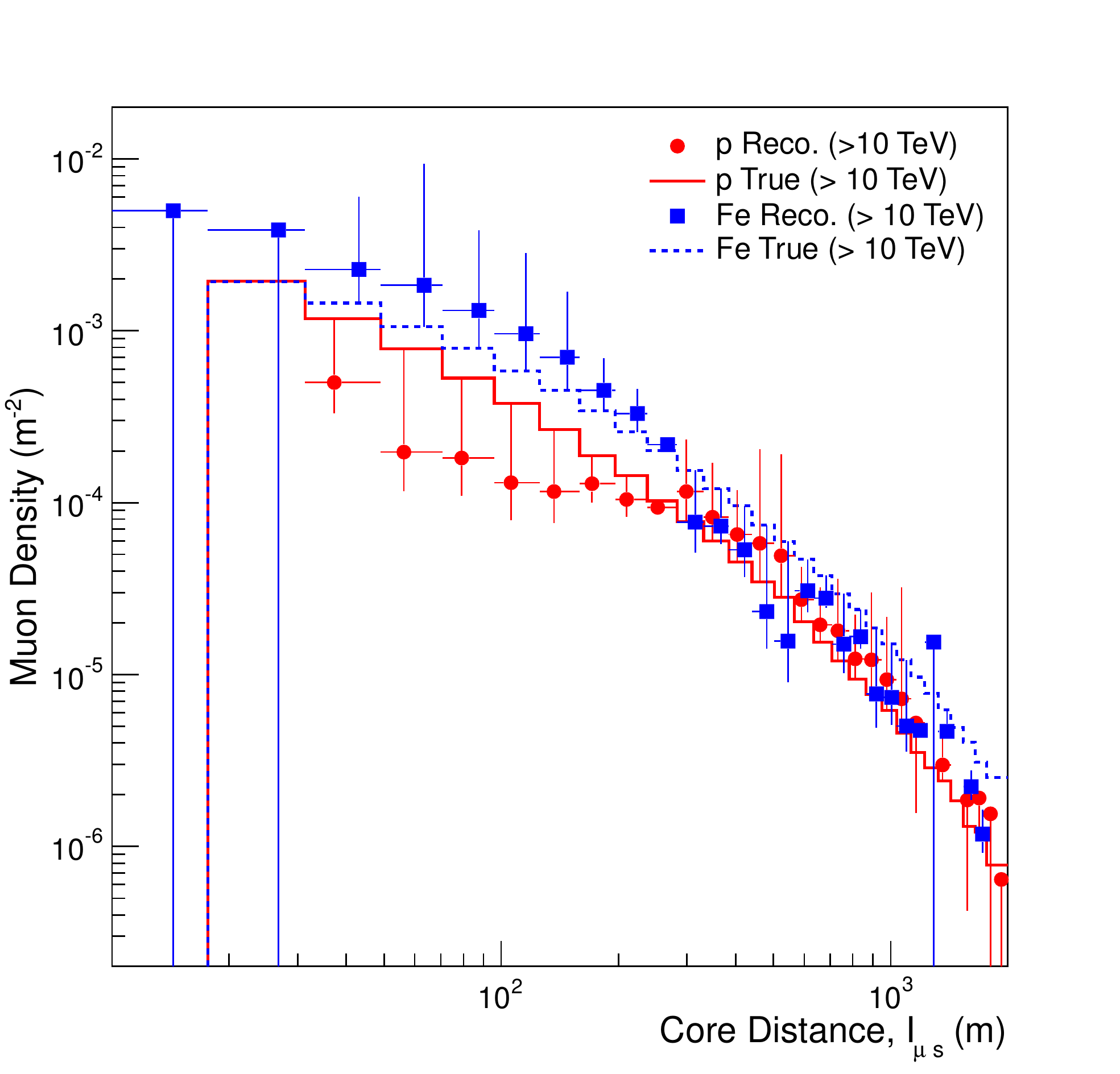}
\caption{Reconstructed muon lateral distribution at ground (points) using effective area to a mixed proton and iron composition, for both proton (red circles) and iron (blue squares) primaries. Shown for QGSJetII-04 only, with lines showing the true $\rho_\mu$ distribution. Error bars show the most pessimistic uncertainties of the method. Left: for 2-10\,TeV showers. Right: for 10-150\,TeV showers. }
\label{fig:recoldf}
\end{figure*}

In the reconstruction, equation \eqref{eq:recomu} is used to find $\rho^*_\mu$; however, as it is not possible in practice for IACTs to separate proton and iron showers with perfect efficiency, we construct an effective area to a mixed composition of equal parts proton and iron as $0.5\times(A^p_{\mathrm{eff}}+A^{Fe}_{\mathrm{eff}})$, using this to reconstruct the muon lateral distribution at ground. The effective areas due to proton and iron were always within a factor 10 of each other, reducing to a ratio of 1:1 at core distances of a few hundred metres where the true distributions cross (as seen in Figure \ref{fig:truemuldf}). This is shown for two energy ranges in Figure \ref{fig:recoldf}. The effective area histograms are somewhat influenced by low statistics at large core distances; smoothing of the histograms was found not to improve the situation. 

The form of the effective area histograms is heavily dependent on the array geometry, with a complex shape exhibiting distinct changes of slope at around $\sim\,100$\,m ($2-10$\,TeV) and $\sim\,700$\,m ($> 10$\,TeV) core distance respectively. A best-fit parameterisation of the effective area histograms with an arbitrary functional form was found to degrade the reconstruction. We therefore use the effective area histograms obtained directly from the simulations as outlined above. 

As an indication of the most pessimistic uncertainties provided by this approach, error bars in Figure \ref{fig:recoldf} show the differences with respect to correcting each sample (of purely proton or purely iron initiated EAS) by the effective area to proton or iron showers in the respective energy range, such that using the effective area to iron showers $A^{Fe}_{\mathrm{eff}}$ for a purely proton initiated sample is maximally pessimistic. 

Figure \ref{fig:recoldf} shows that differences between primaries may be measurable within specific primary energy and core distance ranges; for $2-10$\,TeV showers some separation power at core distances beyond $\sim\,300$\,m is seen, whilst for showers above $10$\,TeV separation power is seen for core distances in the range $\sim\,40-200$\,m (shown for QGSJetII-04 only). Nevertheless, given the good agreement of the true muon lateral distribution curves for 10\,TeV proton showers in Figure \ref{fig:truemuldf}, it is unlikely that differences between hadronic interaction models can be distinguished.

\subsection{Muon production point: slant height}
\label{sec:muonprodh}

\noindent 
In addition to the muon lateral distribution, another quantity that can provide useful information to hadronic interaction models is the muon slant height. Here we use `slant height' to refer to the path length $Z_\mu$ from the muon production point along the shower axis to the telescope, rather than vertical height above ground (see Figure \ref{fig:mugeom}). %

Previous measurements of the muon production height and muon atmospheric production depth have been made by the Kascade-Grande experiment and Pierre Auger Observatory respectively \cite{Kascade11,Auger14maxdepth}.
With the directions and impact positions of both the muon and shower known, the point of closest approach between the muon track and the shower axis in 3D space can be found. The location of this point along the shower axis gives $Z_\mu$, where offsets of the muon production point from the shower axis are neglected.

\begin{figure}[ht]
\begin{center}
\includegraphics[width=\columnwidth]{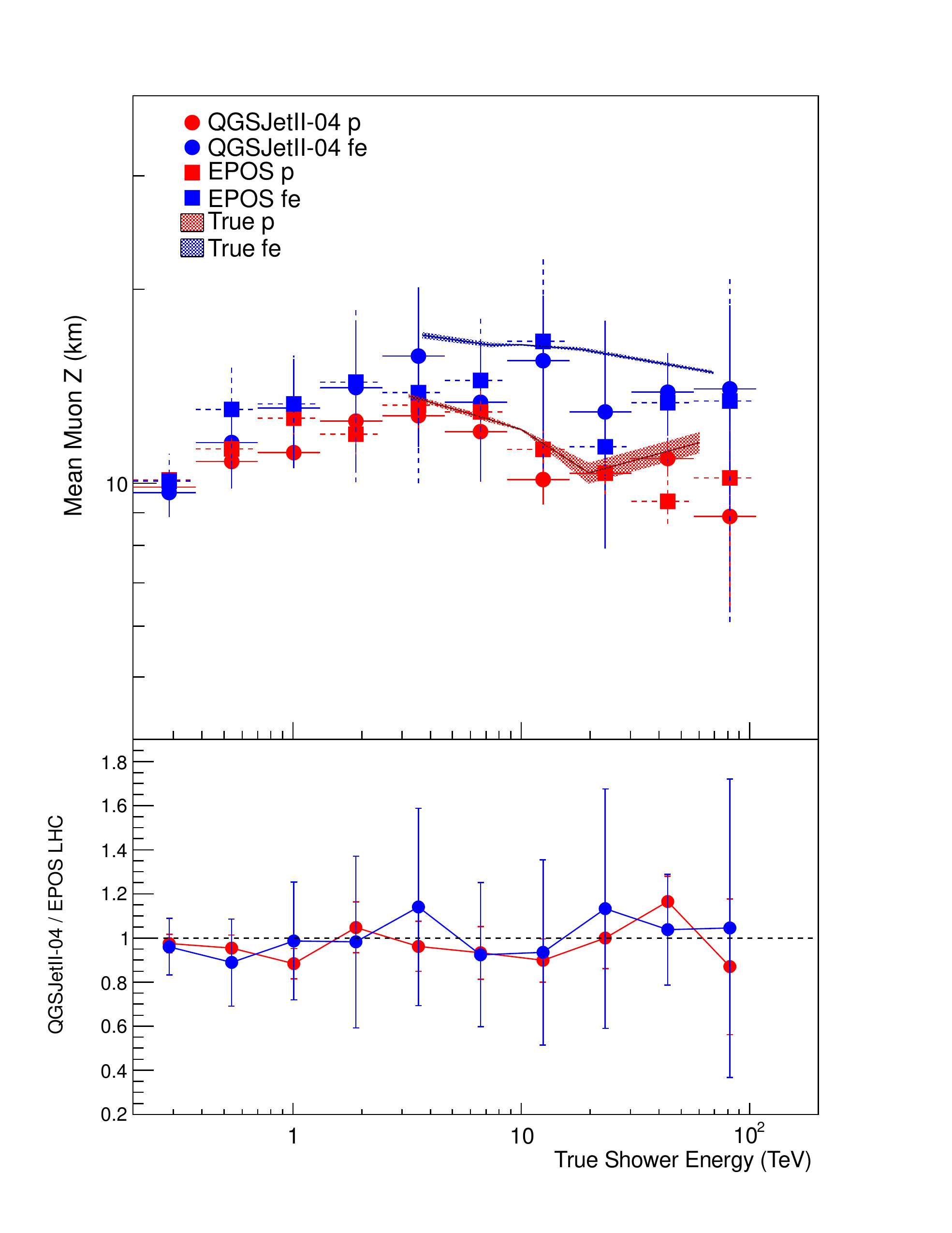}%
\caption{Comparison of the reconstructed muon slant height from proton and iron showers with primary energy $E>2\,$TeV using QGSJetII-04 and EPOS\,LHC. Error bars shown are errors on the mean values with true slant heights shown as bands for comparison. Top: A systematically higher slant height is reconstructed for iron showers than for proton. Bottom: The ratio of reconstructed height between QGSJetII-04 and EPOS\,LHC shows that the two models are in agreement within the uncertainties. }
\label{fig:recoprodheightquad}
\end{center}
\end{figure}

Figure \ref{fig:recoprodheightquad} shows the reconstructed muon slant height as a function of true shower energy from simulated proton and iron showers from two different interaction models. The reconstructed muon slant height for iron showers is marginally systematically higher than that for proton showers over most of the energy range. This would be in agreement with the known shallower depth of shower maximum for iron primaries with respect to proton primaries \citep{KampertUnger12}. However, the reconstructed slant heights from proton and iron primaries are mostly compatible. 
The ratio of reconstructed slant height between QGSJetII-04 and EPOS\,LHC shows that there is no discernible difference between hadronic interaction models in this parameter. 

The muon slant height values, of $\sim\,10-20$\,km are roughly in line with expectation from the true values, with the average for iron showers being slightly higher than that of proton showers. This is expected as the EAS due to heavier primary CRs develop earlier in the atmosphere, such that the height of maximum development is greater and consequently muons are produced on average at greater slant heights.
The reconstructed muon slant height also underestimates the true slant height slightly, consistent with the aforementioned effect of a systematic reduction in reconstructed muon slant height due to neglecting the transverse momentum of parent particles within the shower \citep{Kascade11}. 

Nevertheless, the obtained iron and proton slant heights are broadly compatible within errors. There are indications for changes with energy that remain to be confirmed, although a decrease in production altitude with increasing energy may be expected due to the increasing atmospheric penetration of the primary particle. 

\section{Prospects for the Cherenkov telescope array }
\label{sec:cta}

\noindent The collection area for muons with current generation IACT arrays such as H.E.S.S. is rather low. In comparison to particle shower experiments, there is usually only a single muon detected for each air shower event. Many more muons may contribute to the total light detected by IACTs, as the physical collection area is rather large; however, good muon reconstruction is limited by the need for clean muon events. This results in poor resolution over a small amount of data. 

The resolution may be improved by increased exposure, tighter selection cuts (affordable with a larger data sample), and by an increased collection area. An increased exposure beyond the simulated sample is readily available with the data sets available to current generation IACT arrays, where, in principle, all data taken over the entire operational history may be used. 
An increased collection area, however, would be available with a future facility such as the Cherenkov Telescope Array (CTA). This would also improve the lateral distribution measurement, such that measurements out to larger core distances may be possible without becoming statistics limited.

One obvious way to improve the sensitivity of IACTs to muon lateral distribution measurements is to add more telescopes or use more densely packed telescope array layouts; the forthcoming CTA will be able to improve in both these aspects \citep{Acharya13}.
Comprised of two site locations and three telescope sizes, there is potential for improved measurements of the muon lateral distribution using multiple medium and large sized telescopes, for which both muon and shower can be well reconstructed. 

With a physical array size spread over up to 3~km, as well as high data rates in comparison to those of UHECR experiments, there is great potential for CTA to be able to make the most stringent measurements of muon slant height and lateral distribution from EAS in the $\sim\,10^{11}$eV to $10^{14}$eV energy range.

\begin{figure*}
\centering
\begin{overpic}[width=0.95\columnwidth]{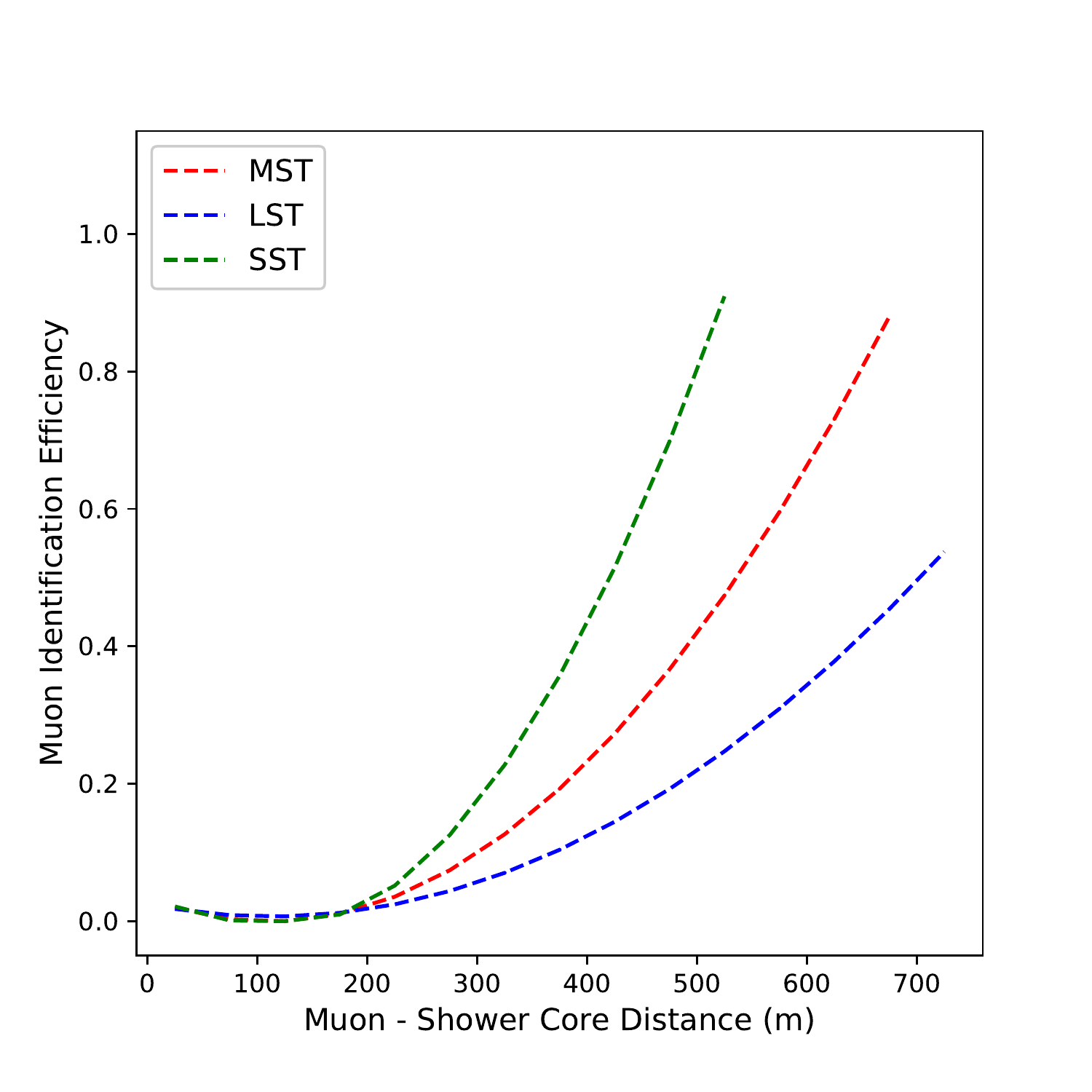}%
\put(75,75){\large{\textcolor{black}{\textmd{\textnormal{CTA}}}}}
\end{overpic}
\hspace{2mm}
\begin{overpic}[width=0.95\columnwidth]{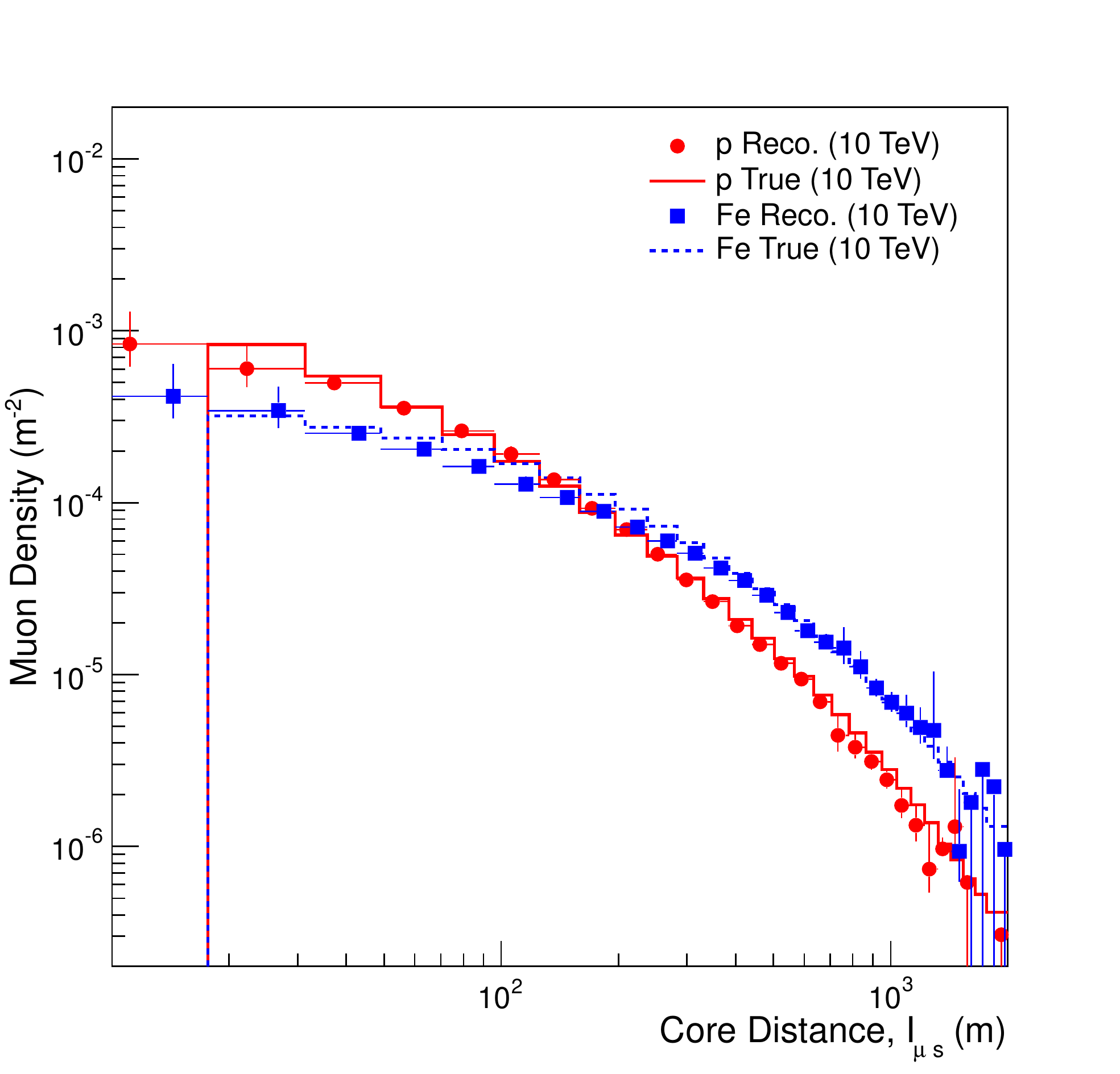}%
\put(15,75){\large{\textcolor{black}{\textmd{\textnormal{CTA}}}}}
\end{overpic}
\caption{Left: Polynomial fits used as lookups for the muon identification efficiency in a model of CTA. Right: Optimistic model for the performance of CTA in recovering the muon LDF, shown for proton and iron showers at 10\,TeV. }
\label{fig:toycta}
\end{figure*}

To test how CTA may perform in recovering the muon LDF, a simple model was constructed using the same set of simulations at 10\,TeV with full particle tracking. The layout of the Southern CTA array, comprising 100 telescopes, was used in place of the HESS telescopes with the position of the muons at ground level used to determine which muons were incident on which telescopes as before. This provided a plausible starting point for the total muon hits for CTA. To determine the fraction of muons that were detected, the muon identification efficiency curves shown in Figures \ref{fig:coredisthitcut} and \ref{fig:angdisthitcut} were parameterised by simple polynomial functions. We assumed that the large telescopes of CTA would behave similarly to CT5 of the HESS array and the medium sized telescopes would behave as the CT1-4 of the HESS array. For the small telescopes of CTA, we derived a muon identification efficiency curve assuming that the performance ratio between the small and medium CTA telescopes is the same as the performance ratio between CT1-4 and CT5 of HESS. To illustrate this, the curves used for the muon identification efficiency as a function of the distance between muon impact position and shower core ($I_{\mu s}$) are shown in Figure \ref{fig:toycta}. The muon identification efficiency as a function of angular distance to the telescope optical axis was determined similarly. 

From the fully tracked simulated EAS, both the true muon impact - shower core distance and angular distance of the muon to the telescope axis could be found. Using these parameterisations, the muon identification efficiency for a given muon hit was inferred. To decide which muons were kept in the analysis, a uniform random number generator was used; if the product of the angular and core distance identification efficiencies was larger than the random number, then the muon was kept. Additionally, to simulate the reconstructed muon - shower core impact distance, the true distance was changed by an amount determined from a random number drawn from a Gaussian distribution with a width of 80\,m, corresponding to the most conservative 68\% containment core resolution found (see section \ref{sec:mushreco}). The selected muons and scattered core distances were used to fill a plausible muon LDF histogram as seen by CTA, shown in Figure \ref{fig:toycta}.

The effective areas of CTA to muons from proton and iron showers at 10\,TeV were determined following the approach described in section \ref{sec:muonldf}, with the reconstructed points for a mixed composition and the error bars showing reconstruction based on a pure iron or a pure proton sample. The reconstructed LDF shown in Figure \ref{fig:toycta} agrees very closely with the true distributions, although larger errors can be seen towards the lowest and highest core distances. It should be noted that we did not apply an additional potential reduction in statistics that could arise from the cut selection on muon ring image quality as shown in Figure \ref{fig:muontp}, nor on the number of triggering telescopes; with the assumption that for the majority of events falling within 1\,km of the centre of the CTA array, multiple telescopes will almost always be triggered by a 10\,TeV hadronic EAS. Additionally, monoenergetic EAS were used; the distribution may degrade for a wider energy range. 

Nevertheless, Figure \ref{fig:toycta} may be considered a conservative estimate of the CTA performance; the increased field of view of CTA telescopes (over the H.E.S.S. telescopes used in this study) and potential improvements in muon identification algorithms have not been considered.

\section{Discussion}
\label{sec:discuss}

In addition to the event selection cuts currently applied in this analysis, one may consider a cut on which telescopes are used in the shower reconstruction; single telescopes exhibiting a large deviation in Hillas parameters with respect to other telescopes in the array could be removed, such that the shower reconstruction is performed with a subset of the triggered telescopes, yet to a higher level of accuracy. Such an approach was trialed in this study, but found not to make a significant difference above the event selection cuts already applied. 
A significant improvement in all measurements could be achieved with improved hadronic shower reconstruction, including energy and core resolution, as well as primary particle identification. Improvements in shower classification beyond current algorithms are out of the scope of this study.

The effective area of an IACT array to muons is given by the area of the telescope mirror dishes at a given distance to the shower core. This therefore varies on an event by event basis, with muons only detected when passing directly through the telescope mirror dish. Additionally, for each telescope a cut on the muon impact distance to the telescope centre is made to ensure good quality reconstruction of the muon image. 

Further improvements in the muon identification efficiency and muon purity could be made by the implementation of more sophisticated ring identification methods, such as by a Hough transform, machine learning or even citizen science approaches, helping to take the intrinsic systematic uncertainties arising from variation in hadronic air shower development into account \cite{Tyler13,Bird18,Feng17}. With a single telescope trigger multiplicity the total number of muon events seen can be significantly increased, albeit without detection of the associated shower \cite{Bolz04}. Additionally, the possibility of separating a muon ring from the hadronic shower in a single telescope in order to use both parts of the image in the reconstruction could be explored, such as by a joint fit of muon ring and Hillas ellipse, machine learning approaches to identify and separate the different components, or by using separate time integration windows for the muon part and the hadronic part of the Cherenkov image. %

With IACTs, it is possible to identify showers from heavier iron-like primaries using a Direct Cherenkov Light (DCL) signal \citep{Kieda01,BuhlerIron07}. To this end, with a good shower classification it may be possible to reconstruct the muon production height and to generate muon lateral distribution functions based on data using both proton-like and iron-like data samples, thereby enabling checks for compatibility of the muon content in both proton and iron showers with hadronic interaction models. 

To estimate the time needed for such a measurement with CTA, we assume that a fraction of $\sim\,1\%$ of events above 10\,TeV are suitable for muon and shower reconstruction, based on Figure \ref{fig:rateenergy}. Events containing DCL suitable for analysis occur at a rate of $\sim\,3\%$ of all events at high energies \cite{BuhlerIron07,VeritasIron18}. For a projected CTA array trigger rate of $\sim\,10$\,kHz, this yields approximately $1\% \times 3\% \times 10$\,kHz $\sim$\,4 events per hour \cite{PazArribas12}. If we assume that all good data taken at zenith angles $\lesssim 40^\circ$ can be used for $\sim\,700$ hours of observing time per year, then similar statistics to Figure \ref{fig:recoldf} ($\sim\,500$ events) can be reached after just a few months of full array operation. Even taking zenith angle binning into account, a reasonable measurement with CTA should be possible within 5 years of data as a conservative estimate.

Use of parameters describing the shower shape to distinguish between different primaries may enable separation without the need for DCL identification, reducing the time necessary for a measurement by increasing the available statistics, albeit with reduced event purity \citep{Ohishi17}.

Alternatively, the muon lateral distribution obtained from data may be used as a measure of the cosmic ray composition at TeV energies; data showing better agreement with proton or iron simulations could indicate whether the detected EAS are predominantly light or heavy. 

It is important to note that the energy reconstruction is based on that for $\gamma$-rays, for which the electromagnetic part of the shower produces the Cherenkov light. In the case of hadronic showers, more energy is lost to other hadrons and neutrinos, as well as to muons, such that the reconstructed shower energy (based only on the electromagnetic shower component) will typically underestimate the true energy. %
The correlation of reconstructed energy with true shower energy was found not to exceed $\sim\,30\%$ with the most stringent cuts applied for the simple shower reconstruction used in this study.
Whilst the use of a cut on the deviation in Hillas parameters improves the performance of the reconstruction by restricting the analysis to $\gamma$-like showers, this also introduces a strong bias in the sample of hadronic showers used.

Given the results presented in section \ref{sec:results}, it may not be possible to make an absolute measurement of the true muon density in a model-independent manner. Similarly, absolute measurements of the muon production height may not be possible without further considerable method development. Nevertheless, relative measurements may still be able to provide useful information and place important constraints on the compatibility of hadronic interaction models with data \citep{TAmuon}.

Cosmic ray muon data is obtained essentially `for free' by IACTs, being recorded amongst the $\gamma$-ray showers that are the main focus of IACT arrays. As muons will be kept for calibration purposes, one only needs to ensure that the associated shower images from other telescopes are also retained, and that the total number of triggered events above a defined threshold in image size and telescope multiplicity counted, in order to make measurements that could provide significant contributions to the cosmic ray community and hadronic interaction model development. 
However, keeping full event information for all shower events is neither a necessary nor a viable option for CTA, given the vast data rates expected under normal operation of up to $\sim\,5.4$\,Gbps raw data \citep{Lamanna15}.

Future IACT implementations may also make further use of timing information in shower recording. This may help not only in $\gamma$-ray reconstruction and as a $\gamma$/hadron discriminator, but also in reconstructing more accurately the muon production height \cite{Auger14muondepth}. As muon signals typically peak in advance of the shower, an optimised signal integration window could reduce the level of muon image contamination by signal from the parent shower, improving the muon reconstruction and clean muon rate \citep{Bernlohr13}. 

Further studies could include investigating a potential zenith angle or energy dependence of the shape of the muon lateral distribution at ground; which would be expected due to the angle of the incident shower.
For a measurement on data, binning of the effective area to muons in both energy and zenith angle would be necessary for reconstruction of the lateral distribution.

These results are encouraging that such measurements can be performed with IACTs. Relative measurements of the effective muon density are possible regardless of the ring identification method used, as the rate of clean, detectable muons with any single algorithm and set of cuts is a constant fraction of the total muon flux.

\section{Conclusion}
\label{sec:conclusion}

\noindent IACTs offer a complementary approach to verifying hadronic interaction models using CR data in the TeV range. Currently there are few measurements performed at these energies, yet this range is ideal for constraining whether the discrepancies between EAS and accelerator measurements originate in the extrapolation of models towards higher energies or a more fundamental understanding of the physics of air shower development.

Whilst hadronic air showers typically form a background signal for IACT experiments, we demonstrate that there are important, measureable parameters, such as the effective muon density of hadronic air showers, that can also be constrained with IACTs. 

As the community gears towards the future CTA observatory, we consider it important that such additional scientific cases are considered already in the design of such a facility, enabling measurements significant for the cosmic ray community to be feasible in the more long-term future.

With a higher telescope density, such as in the central part of the Southern CTA array, the effective area to muons will be much increased over that currently possible with IACT arrays such as HESS \citep{Acharya13,Bernlohr13}. Nevertheless, a first measurement by the current generation of IACT experiments would be a welcome input to the development of hadronic interaction models.  

\vspace{0.5cm}
\noindent \textbf{Acknowledgements}
\newline
We would like to thank W.\,Hofmann, J.\,Hinton and J.\,Knapp for useful discussions, as well as K.\,Bernl\"{o}hr for assistance with the simulations. We thank the H.E.S.S. collaboration for permission to use the HESS Analysis Package software framework. 

\bibliographystyle{abbrvnat}
\bibliography{RecoMuonRefs.bib}

\end{document}